\documentclass[fleqn,usenatbib]{mnras}

\usepackage[T1]{fontenc}

\DeclareRobustCommand{\VAN}[3]{#2}
\let\VANthebibliography\thebibliography
\def\thebibliography{\DeclareRobustCommand{\VAN}[3]{##3}\VANthebibliography}

\usepackage{graphicx}   
\usepackage{amsmath}    
\usepackage{amssymb}    
\usepackage{pdflscape}
\usepackage{tablefootnote}
\usepackage{txfonts}

\title[Monte-Carlo simulations of rotational distributions]{Monte-Carlo
simulations of evolving rotational distributions of low-mass stars
in young open clusters. Testing the influence of initial conditions.}

\author[M. J. Vasconcelos et al.]{M. J. Vasconcelos,$^{1}$
\thanks{Contact e-mail: \href{mailto:mjvasc@uesc.br}{mjvasc@uesc.br}}
J. Bouvier,$^{2}$
F. Gallet,$^{2}$
and E. A. Luz Filho$^{1}$ 
\\
$^{1}$LATO–DCET, Universidade Estadual de Santa Cruz, UESC, Rodovia
Jorge Amado, km 16, Ilhéus, 45662-900, Brazil \\
$^{2}$Univ. Grenoble Alpes, CNRS, IPAG, Grenoble, 38000, France}

\date{Accepted 2021 November 17. Received 2021 November 10; in
original form 2020 December 03}

\pubyear{2020}

\begin{document}
\label{firstpage}
\pagerange{\pageref{firstpage}--\pageref{lastpage}}
\maketitle

\begin{abstract}
The rotational evolution of a young stellar population can give
informations about the rotation pattern of more evolved clusters.
Combined with rotational period values of thousands of young stars
and theoretical propositions about the redistribution and loss of
stellar angular momentum, it allows us to trace the rotational
history of stars according to their mass. We want to investigate
how internal and environmental changes on single stars can change
the rotational evolution of a young stellar population. We run Monte
Carlo simulations of a young cluster composed by solar mass stars
of 0.5, 0.8 and 1.0 M$_\odot$ from 1 to 550 Myr taking into account
observational and theoretical parameters.  In order to compare our
results with the observations we run Kolmogorov-Smirnov tests. Our
standard model is able to reproduce some clusters younger than h
Per and marginally M37, which is 550 Myr old. Varying the disk
fraction or the initial period distribution did not improve the
results. However, when we run a model with a finer mass grid the
Pleiades can be also reproduced. Changing the initial mass distribution
to be similar to the empirical ONC mass function also gives good
results. Modeling the evolution of a young synthetic cluster from
pre-main sequence to early main sequence considering physical
mechanisms of extraction and exchange of angular momentum can not
be achieved successfully for all clusters for which we have enough
rotational data. Clusters of about the same age present different
rotational behaviors due perhaps to differences in their initial
conditions.

\end{abstract}


\section{Introduction}

\def\PaperI{\href{http://adsabs.harvard.edu/abs/2015A&A...578A..89V}{Paper~I}}
\def\PaperII{\href{http://adsabs.harvard.edu/abs/2017A&A...600A.116V}{Paper~II}}

Rotation is an important ingredient of stellar evolution. It
influences the stellar dynamo and some parameters such as the stellar
radius and lithium content \citep{2015MNRAS.449.4131S}. Recently,
thousands of light curves from different ground and space missions
like Kepler's K2 and CoRoT provided high quality rotational periods
of young stars in different clusters
\citep{2014prpl.conf..433B,2020AJ....159..273R}.  It becomes
clear that the stellar rotation period depends on age as well as
on the mass of the star. It can also be seen that younger pre
main-sequence (PMS) clusters present a wider spread in periods than
older, main-sequence (MS) clusters. At about 600 Myr, a slow rotator
sequence develops for stars between 0.6 M$_\odot$ and 1.0 M$_\odot$
while lower mass stars continue to present faster rotating stars
\citep{2017ApJ...839...92R}.

It is believed though that the rotation pattern seen in different
clusters can be reproduced evolving the individual stellar angular
velocities \citep{1998A&A...333..629A}. During the PMS phase, the
contraction of the stellar radius and the development of a radiative
core will decrease the moment of inertia resulting in a rotational
acceleration of the star. On the MS, a stellar magnetized wind will,
in turn, carry matter and angular momentum away from the star,
decelerating it.  The star can rotate as a solid body until the
development of a radiative core for stars more massive than 0.3
M$_\odot$. In this case, the core can store angular momentum that
can be released afterwards and alter the angular velocity of the
convective envelope of the star.

Another important ingredient in modeling the rotation of a young
star is the magnetic coupling between an accretion disk and the
star, or disk locking mechanism. During the Hayashi phase, it is
expected that the contraction of the star will increase its angular
velocity. Also, the accretion flow would carry angular momentum to
the star. However, this is not what is observed. PMS stars showing
signs of surrounding accretion disks seem to rotate more slowly
than diskless stars
\citep{1993AJ....106..372E,2004AJ....127.1029R,2007ApJ...671..605C}.
Torques due to the interaction of the stellar magnetic field and
the accretion disk have been proposed as mechanisms that prevent
the star from gaining too much angular momentum.
\citet{1991ApJ...370L..39K} applied this idea in the context of
accreting T Tauri stars after \citet{1979ApJ...234..296G} model for
pulsating X-ray sources. Other mechanisms of angular momentum loss
for young accreting stars have been proposed by several authors:
the X-wind model by \citet{1994ApJ...429..781S}; accretion powered
stellar winds by \citet{2005ApJ...632L.135M}; reconnection X-winds
by \citet{2000MNRAS.312..387F}; magnetospheric ejections by
\citet{2009MNRAS.399.1802R} and \citet{2013A&A...550A..99Z} among
others. In spite of this, angular momentum regulation during the
accretion phase has not been fully incorporated in angular momentum
evolution models.  Usually, this is modeled keeping the stellar
angular velocity constant.  Recently, \citet{2019A&A...632A...6G}
combined the \citet{2013A&A...556A..36G} rotational evolutionary
model with \citet{2013A&A...550A..99Z} results of numerical simulations
of magnetospheric ejections in order to treat self-consistently the
stellar spin rate evolution during the accreting phase.  Their
results require stringent conditions in order to reproduce the 25th,
50th and 90th percentiles of the rotational distributions of solar
type stars in five clusters with ages spanning 1.5 Myr and 13 Myr:
kilogauss stellar magnetic fields, stellar winds with a mass loss
rate of the order of 10\% of the mass accretion rate or lighter
winds (1\% of the mass accretion rate) combined with a disk truncated
beyond the corotation radius which leads the system to a strong
variability that is not seen in the observations.
\citet{2021ApJ...906....4I} performed simulations similar to those
by \citet{2019A&A...632A...6G} exploring a greater parameter space
and taking into account a wind torque related to the accretion
process.  They suggest three parametric torque terms due to accretion,
magnetospheric ejections and a stellar magnetized wind.  Their
simulations show that the spinning up torques from accretion and
magnetospheric ejections dominate over the spinning down one, from
the stellar wind. To achieve a rotational equilibrium configuration
a high wind mass loss rate is required.

\begin{figure*}
    \centering
    \includegraphics[width=\textwidth]{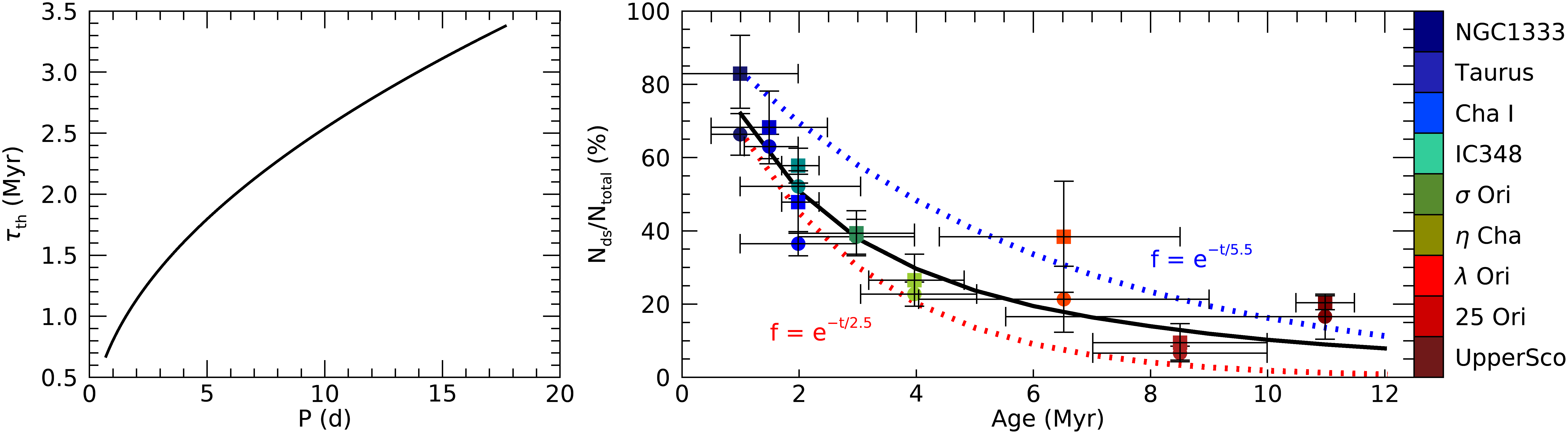}
    \caption{Left panel: the parameter $\tau_\mathrm{th}$ as a
    function of rotational period. Right panel: disk fraction as a
    function of cluster age superimposed on data for nine young
    nearby clusters, associations and groups shown as colored dots
    taken from \citet[][circles]{2014A&A...561A..54R},
    \citet[][squares]{2007ApJ...662.1067H,2008ApJ...686.1195H}.
    The cluster ages were taken from \citet{2014A&A...561A..54R}.
    The dotted lines are exponential decay laws expected from disk
    e-folding times 2.5 Myr (red line) and 5.5 Myr (blue line).
    Both figures are obtained for $\overline{P} = 7.5$ d.}
    \label{tth75}
\end{figure*}

Vasconcelos \& Bouvier (\citeyear{2015A&A...578A..89V}, hereafter
Paper~I) created a synthetic cluster of rotating solar mass stars
(0.3 - 1.0 M$_\odot$). The paper's goal was to reproduce the
rotational period distributions seen in clusters younger than 20
Myr, particularly, the ONC, NGC 2264 and h Per. For this purpose,
the stellar angular velocities were evolved taking into account
only the spin up due to PMS contraction and without considering any
angular momentum loss. In \PaperI, the stellar population was divided
in disk and diskless stars and only the latter had their angular
velocities evolved. This simple approach was successful in reproducing
qualitative aspects of the ONC, NGC 2264 and Cyg OB2 stellar periods,
the relation between disk fraction and mass accretion rate and
rotational period but failed to capture the striking bimodality of
h Per. In Vasconcelos \& Bouvier (\citeyear{2017A&A...600A.116V},
Paper~II) the same kind of simulations were performed but in the
brown dwarf and very low mass regime. The results show that
the rotational behavior of these low mass objects is different from
their higher mass counterparts. In this case, the disk locking is
not so necessary in order to reproduce the observed period
distributions. The conclusion was that these lower mass objects
should rotate faster than solar mass stars.

In this paper, we want to extend \PaperI's simulations to include
angular momentum loss through stellar magnetized winds and the
core-envelope decoupling in order to investigate the rotational
pattern of clusters as old as the 550 Myr old M37. There are some
previous similar works. \citet{2011MNRAS.416..447S} generated a
synthetic population of solar mass stars and evolved it from 2 Myr
to 4.5 Gyr considering disk locking, core-envelope coupling and
angular moment loss by a wind. They focused in obtaining prescriptions
for the core-envelope coupling timescale and optimized parameters
and compared their synthetic populations to clusters of similar
ages. \citet{2017A&A...599A.122G} also constructed a synthetic
population at 5 Myr based on the period distribution of NGC 2362
consisting of 1000 0.7 - 1.1 M$_\odot$ stars with 0.3 d $\leq$ P
$\leq$ 30 d. No disk locking was imposed and a constant value
core-envelope coupling timescale was assumed and also an enhanced
angular momentum loss attached to a Rossby number interval.  According
to the author, this ``catastrophic" braking will generate bimodal
period distributions from 20 Myr to 600 Myr that are in agreement
with the observations. \citet{2016ApJ...833..122C} used YREC stellar
code \citep{2008Ap&SS.316...31D} that solves 1D equations of the
stellar structure taking into account structural changes due to
rotation, the presence of disks and winds and internal angular
momentum transport. They evolved the period distribution of a
population of 0.3 - 0.7 M$_\odot$ stars from 1 Myr to 13 Myr and
then to 550 Myr. Initially they construct a distribution based on
the ONC periods of disk and diskless stars taken from
\citet{2006ApJ...646..297R} data. The distributions are evolved
from 1 Myr to 13 Myr and compared to h Per data. The authors concluded
that they cannot reproduce the fast h Per rotators peak. After,
they evolved a new distribution obtained from h Per data by
\citet{2013A&A...560A..13M} from 13 Myr to 550 Myr. After comparing
their results to the Pleiades
\citep{2010MNRAS.408..475H,2016ApJ...822...81C} and M37
\citep{2009ApJ...691..342H} data they concluded that they can
reproduce the slow rotators but not the fast rotators of these
distributions. \citet{2015A&A...584A..30L} constructed two-zone
models testing different wind braking prescriptions: i) the
\citet{1988ApJ...333..236K} law as modified by \citet{1995ApJ...441..876C};
ii) the \citet{1988ApJ...333..236K} law with the mass dependence
suggested by \citet{2010ApJ...721..675B} and \citet{2010ApJ...722..222B};
iii) the braking law proposed by \citet{2013A&A...556A..36G} and
iv) the braking law by \citet{2015ApJ...799L..23M}. They assume the
same constant rotational period and disk lifetimes for the stars
independent of their mass for all models. They constrain the set
of free parameters necessary to describe the slow rotator sequence
of clusters from 0.1 to 2.5 Gyr for the mass range 0.85 - 1.1
M$_\odot$ using a Monte Carlo Markov Chain method. Their results
show that in order to reproduce the observations the coupling time
scale related to the angular momentum exchange between the stellar
radiative interior and the convective envelope should vary with the
mass. Concerning the wind prescriptions, they conclude that all
models reproduce well the observations but prescription ii) is
better.  In order to reproduce the slow rotator sequence of the
main sequence open clusters Praesepe (700 Myr) and NGC 6811 (1 Gyr),
\citet{2020A&A...636A..76S} used the \citet{2015A&A...584A..30L}
model with parameters adjusted to better reproduce the clusters.
They adopted the wind braking law with \citet{2010ApJ...721..675B}
mass and \citet{1988ApJ...333..236K} $\omega_\mathrm{env}^3$
dependencies (wind prescription ii above).  The rotational coupling
timescale shows a stronger mass dependency compared to what was
obtained by \citet{2015A&A...584A..30L}. They show that this is
essential to capture the rotational behavior of 0.5 - 0.8 M$_\odot$
stars which present almost the same rotational periods in the two
clusters. They conclude that at this mass range, resurfacing of
angular momentum causes an apparent reduction of NGC 6811 spinning
down. In this sense, the evolution of the rotational distribution
of older clusters is not only due to the wind loss but depends also
on internal exchanges of angular momentum.

In this work, we will investigate if we can reproduce the rotational
pattern of some clusters using disk locking, angular momentum loss
by a magnetized wind and internal exchange of angular momentum.
Differently from these earlier works, the disk lifetime will be a
function of the initial rotation period and we will consider the
wind angular momentum loss given by \citet{2013A&A...556A..36G,
2015A&A...577A..98G}. We will experiment different disk timescales
and initial conditions.  The models will be compared to the
observations by means of the statistical Kolmogorov-Smirnov test.
In Section~\ref{model_setup} we present the equations and the
numerical setup of our simulations.  In Section~\ref{M1} we discuss
our standard model, M1.  Variations upon this model will be present
in Section~\ref{difparam} and its subsections. Finally in
Section~\ref{conclusions} we draw our conclusions.

\section{The numerical setup} \label{model_setup}

Our numerical setup is very similar to that of \PaperI. We start
our simulations at 1 Myr creating initial distributions for the
rotational period and the mass accretion rate of the population.
The rotational period distribution is a bimodal truncated Gaussian.
The distribution is limited to 0.7 and 18.5 days. The mass accretion
rate initial distributions are log-normal functions with a mean
equal to the logarithm of $10^{-8}$ (M$_\ast$/M$_\odot)^{1.4}$
M$_\odot$ yr$^{-1}$ and a standard deviation $\sigma = 0.8$. The
mass values considered in this work are equal to 0.5, 0.8 and 1.0
M$_\odot$ The number of stars in each mass bin is calculated according
to the canonical IMF by \citet{2013pss5.book..115K}. Other stellar
mass values and distributions will be investigated in Section~\ref{MF}.
The total number of stars is around 50,000. The mass accretion rate
is evolved at each time step and varies with t$^{-1.5}$
\citep{1998ApJ...495..385H}.

Until 12 Myr, we consider the disk locking hypothesis and the angular
velocity (or period) of disk stars is constant although as can be
seen in Fig.~\ref{tth75}, at this age, most of the stars no longer
have disks. According to the observations, the expected disk lifetime
is shorter than 12 Myr. With our hypothesis stars with long living
disks will not have much time to spin up significantly (see
Fig.~\ref{wevol75b}). The presence or absence of the disk is
determined by the comparison of the stellar mass accretion rate to
a threshold which is given by,

\begin{equation} \label{Macctth}
\dot{\mathrm{M}}_\mathrm{acc,th} = 10^{-8} \left(2.2
\sqrt{\frac{P_0}{\overline{P}}}\right)^{-1.5} (M_\ast/M_\odot)^{1.4}.
\end{equation}

\noindent It depends on the initial stellar period, $P_0$, on
$\overline{P}$, a free parameter, and on stellar mass. In \PaperI\
we have shown that at an age of 2.2 Myr, half of the stars have
lost their disks.  Here, the mass accretion rate threshold depends
on the stellar initial rotating rates as suggested by
\citet{2013A&A...556A..36G,2015A&A...577A..98G}. Slowly rotating
stars, for example, will have smaller mass accretion rate thresholds.
A small $\dot{\mathrm{M}}_\mathrm{acc,th}$ value implies longer
lasting disks, since it will take longer for the mass accretion
rate to reach the threshold. One can define a time scale
$\tau_\mathrm{th}$ given by the expression inside the parentheses
in Eq.~\ref{Macctth}. In the left panel of Fig.~\ref{tth75}, we
show $\tau_\mathrm{th}$ as a function of the rotational period
considering $\overline{P} = 7.5$ d. The longer $\overline{P}$ the
shorter $\tau_\mathrm{th}$. One can note that the maximum
$\tau_\mathrm{th}$ value obtained for $\overline{P} = 7.5$ d is
about 3.5 Myr. When analyzing disk fractions as a function of cluster
age (right panel of Fig.~\ref{tth75}), we note that this $\overline{P}$
value gives a disk fraction which is in agreement with most of the
observed values and between two exponential curves with e-folding
times equal to 2.5 Myr and 5.5 Myr. Equation~(\ref{Macctth}) for
the parameter $\tau_\mathrm{th}$ provides a range of mass accretion
rate thresholds instead of just one value per mass.  If
$\dot{M}_\mathrm{acc}(t) \le
\dot{M}_\mathrm{acc,th}(M_\ast,P_0,\overline{P})$ the star loses
its disk. Then, the angular velocity is evolved. After 12 Myr, all
stars will be considered diskless, even if their mass accretion
rates are above $\dot{M}_\mathrm{acc,th}$.

PMS stars can lose angular momentum through a stellar magnetic wind.
For the mass interval considered in this work, the stars will develop
a radiative core. In this case, it is expected some kind of exchange
of angular momentum between the core and the envelope.  There is
no definitive mechanism to explain this exchange but the best ones
are related to internal gravity waves, hydrodynamical instabilities
and magnetic fields \citep[for a short review, see][]{2014prpl.conf..433B}.
We implement the so-called double zone model in which both the core
and the envelope are considered solid bodies that interact at a
rate given by a free parameter, the core-envelope coupling timescale,
$\tau_{ce}$ \citep{1991ApJ...376..204M,2007MNRAS.377..741I}. The greater
$\tau_{ce}$, the weaker the core-envelope interaction. It means
that a smaller amount of angular momentum will be exchanged by the
core and the envelope. The equations for the evolution of the
envelope and core angular velocities ($\omega_{env}$ and $\omega_{core}$,
respectively) for diskless stars are given by,

\begin{align} \label{wenv}
\begin{split}
\frac{d\omega_{env}}{dt} = & \frac{1}{I_{env}}\frac{\Delta J}{\tau_{ce}} - 
\frac{2}{3}\frac{R^2_{core}}{I_{env}}\omega_{env}\frac{dM_{core}}{dt}
- \frac{\omega_{env}}{I_{env}}\frac{dI_{env}}{dt} \\
\\
-  & \frac{w_{env}}{J_{env}}\dot{J}_\mathrm{wind},
\end{split}
\end{align}

\begin{align} \label{wcore}
\begin{split}
\frac{d\omega_{core}}{dt} = & - \frac{1}{I_{core}}\frac{\Delta J}{\tau_{ce}} + 
\frac{2}{3}\frac{R^2_{core}}{I_{core}}\omega_{env}\frac{dM_{core}}{dt} \\
\\
- & \frac{\omega_{core}}{I_{core}}\frac{dI_{core}}{dt},
\end{split}
\end{align}
with,
\begin{equation} \label{deltaJ}
\Delta J = \frac{I_{env} J_{core}-I_{core}J_{env}}{I_{env} + I_{core}},
\end{equation}

\begin{table}
    \centering
    \caption{Core-envelope coupling timescales}
    \begin{tabular}{|c|ccc|}
    \hline \hline
      Mass (M$_\odot$) & Slow & Medium & Fast \\
                       & \multicolumn{3}{|c|}{Rotators (Myr)} \\
      \hline
      0.5 & 500 & 300 & 150 \\
      0.8 &  80 &  80 &  15 \\
      1.0 &  30 &  28 &  10 \\
    \hline
    \end{tabular}
    \label{initcond}
\end{table}

\begin{equation}
I_\mathrm{core,env} = k_\mathrm{core,env}^2 M_\ast R_\ast^2,
\end{equation}

\noindent In the equations above, the symbols have their usual
meanings, namely, the moments of inertia $I$ (for the core and
envelope, $I_{core}$ and $I_{env}$, respectively), angular velocity
$\omega$ (for the core and envelope, $\omega_{core}$ and $\omega_{env}$),
the core radius ($R_{core}$), the angular momentum $J$ (for the
envelope and the core, $J_{env}$ and $J_{core}$), and the mass $M$
(for the star and for the core, $M_{\ast}$ and $M_{core}$).
$\dot{J}_\mathrm{wind}$ is the torque due to the magnetized stellar
wind and $k_\mathrm{core,env}$ is the gyration radius for the core
or the envelope. They were calculated by \citet{1998A&A...337..403B}
using the expressions:

\begin{equation}
k_\mathrm{core,env} = \sqrt{\frac{I^\mathrm{{mod}}_\mathrm{{core,env}}}{M_\ast R_\ast^2}},
\end{equation}

with,

\begin{equation}
I^\mathrm{{mod}}_\mathrm{{core,env}} = \frac{2}{3} \int r^2 dm.
\end{equation}
The integral in the equation above is calculated over the mass of
the convective regions and the regions below. The variable $dm$ is
the mass of a thin shell of radius $r$ inside a spherical star.

Since we are considering the core and the envelope as bodies in
solid rotation, the equations of angular velocity evolution have
terms related to angular momentum variation (torques) and terms
related to the stellar contraction (i.e. to the evolution of the
moment of inertia).  These last ones are the 3$^{rd}$ terms in
equation~(\ref{wenv}) and equation~(\ref{wcore}). In the PMS, due
to contraction, the star is spun up since its total moment of inertia
decreases. When the radiative core develops, the moment of inertia
of the core increases but the moment of inertia of the envelope
decreases and so these terms contribute to decrease the core angular
velocity $\omega_\mathrm{core}$ but to increase $\omega_\mathrm{env}$.
This imbalance stabilizes when the star reaches the main sequence.
The first terms in equation~(\ref{wenv}) and equation~(\ref{wcore})
are torque terms related to the core-envelope exchange of angular
momentum \citep{1991ApJ...376..204M}. They can be positive or
negative depending on the relation between $\omega_\mathrm{core}$
and $\omega_\mathrm{env}$ and they always act to establish uniform
internal rotation. Due to the mass and radius increase of the core,
initially $\omega_\mathrm{core} >\omega_\mathrm{env}$ and so there
is a transfer of angular momentum to the envelope which will occur
over the $\tau_\mathrm{ce}$ timescale. The second terms are torque
terms due to the mass increase of the radiative core
\citep{1998A&A...333..629A}.  Finally, the last term in
equation~(\ref{wenv}) is the torque due to the the magnetized stellar
wind. Stellar parameters are taken from \citet{1998A&A...337..403B}'s
models.This stellar evolutionary model neglects the effects of
accretion \citep[see, e.g.,][]{2010A&A...521A..44B} and rotation
\citep[e.g.,][]{2016A&A...586A..96L, 2019A&A...631A..77A}. All terms
related to the development of the core have no effect until around
3 - 4 Myr for 0.8 - 1 M$_\odot$ stars and later, around 13 Myr, for
0.5 M$_\odot$ stars. Our models use the angular momentum loss due
to a magnetized wind given by \citet{2015A&A...577A..98G} in the
form,

\begin{equation}
\Gamma_\mathrm{wind} = \omega_\ast \dot{M}_\mathrm{wind}
r_\mathrm{A}^2,
\end{equation}
where $\omega_\ast$ is the stellar angular velocity,
$\dot{M}_\mathrm{wind}$ is the mass loss rate given by the numerical
simulations of \citet{2011ApJ...741...54C} and modified by
\citet{2013A&A...556A..36G} and $r_\mathrm{A}$ is the average value
of the Alfvén radius given by \citet{2012ApJ...754L..26M}.

We consider three models, M1, M2 and M3. Models M1 and M2 share the
same initial period distributions for disk and diskless stars.  For
disk stars, the distribution has a mean equal to $\overline{P}_\mathrm{d}
= 7.0$ days and a dispersion of $\sigma_\mathrm{d} = 3.0$ days. For
diskless stars, we adopt a distribution with $\overline{P}_\mathrm{dl}
= 2.0$ days and $\sigma_\mathrm{dl} = 4.0$ days.  All models use
the same core-envelope coupling timescale $\tau_\mathrm{ce}$ values
\citep{2015A&A...577A..98G}, which are shown in Table~\ref{initcond}.
Models M1 and M2 have different $\overline{P}$, equal to 7.5 days
and 2.0 days, respectively, which normalizes the mass accretion
rate threshold (equation~\ref{Macctth}).  Model M3 begins at 13 Myr
with a period distribution taken from the observational period
distribution of h Per \citep{2013A&A...560A..13M}.  We also examine
the effect that other mass values and distributions have on
model M1 in Section~\ref{MF}. Torques due to the magnetized wind
were calculated for a grid of initial periods from 0.05 to 45 days
using \citet{2015A&A...577A..98G} prescription. The values for the
actual period distributions were then obtained by interpolation.
The values necessary to calculate the other torque terms were taken
or calculated from \citet{1998A&A...337..403B} stellar models for
each one of the masses.  The core-envelope coupling timescales for
models M1, M2 and M3 have the same values as those of
\citet{2015A&A...577A..98G}, considering their definition of slow,
median and fast rotators. The corresponding values for the intermediate
masses used in section \ref{MF} have all been obtained by interpolation,
except the stellar parameters for which we have the values from
\citet{1998A&A...337..403B} tables.

\begin{table*}
\renewcommand{\arraystretch}{1.5}
  \centering
  \caption{Selection criteria for the observational samples.}
    \begin{tabular}{ccccccc}
      \hline \hline
      Cluster & Age (Myr) & Age Ref. & Mass/Spt/Color index range &
      Disk?$^a$ & N$^b$ & Ref. \\
      \hline
      ONC & 2 & \citet{2008MNRAS.386..261M} & K4.5 - M0 & [3.6]
      - [8.0] $>$ 0.7 & 68 & \citet{2007ApJ...671..605C} \\
      ONC & 2 & " & K4.5 - M0 & Class/Accreting & 45 &
      \citet{2014MNRAS.444.1157D} \\
      Cyg OB2$^c$ & 5 & \citet{2010ApJ...713..871W} & 0.5 - 1.0
      M$_\odot$ & 0 - 1 & 258 & Roquette et al. (2017) \\
      NGC 2362 & 4-5 & \citet{2008MNRAS.386..261M} & 0.5 - 1.0 M$_\odot$
      & - & 88 & \citet{2008MNRAS.384..675I} \\
      U Sco & 11 & \citet{2012ApJ...746..154P} & $2.0 \le (V - K_s)_0
      \leq 4.1$ & - & 144 & \citet{2018AJ....155..196R} \\
      h Per & 13 & \citet{2008MNRAS.386..261M} & 0.5 - 1.0 M$_\odot$
      & - & 219 & Moraux et al. (2013) \\
      Pleiades & 125 & \citet{1998ApJ...499L.199S} & 0.5 - 1.0
      M$_\odot$ & - & 184 & Rebull et al. (2016), Lodieu et al.
      (2019) \\
      M37 & 550 & \citet{2008ApJ...675.1233H} & $0.76 \le (V - I)
      \le 2.0$ & - & 366 & \citet{2009ApJ...691..342H} \\
      \hline
      \multicolumn{3}{c}{$^a$ Disk selection criteria as found in
      the respective references.~~ } & & &   \\
      \multicolumn{3}{c}{$^b$ Number of stars after the application
      of the selection criteria.} & & &  \\
       \multicolumn{3}{c}{$^c$ Stars with periods greater than 2
       days. See text for details.} & & & \\
    \end{tabular}
    \label{tabcritobs}
\end{table*}

\begin{figure}
  \centering
  \includegraphics[width=8cm]{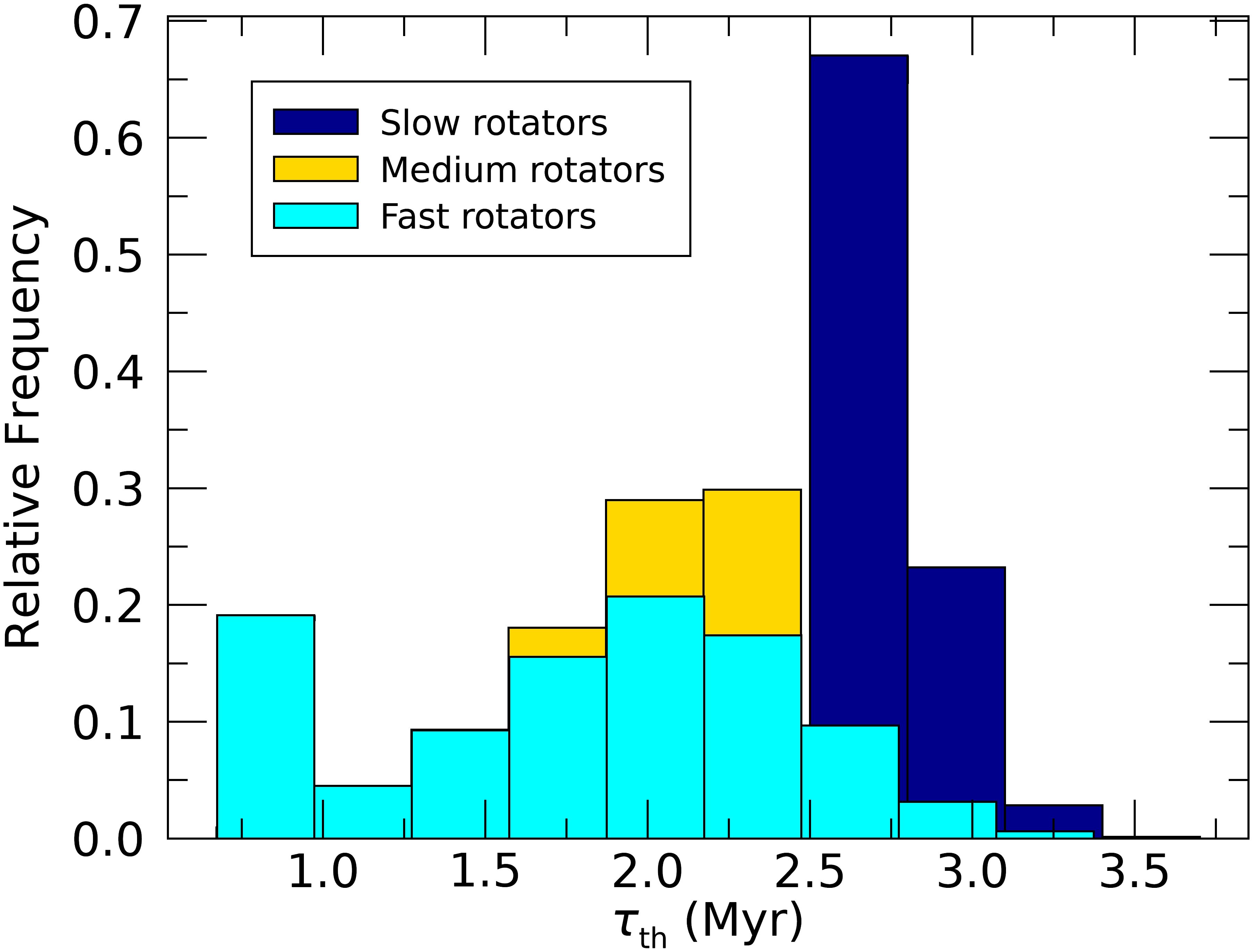}
  \includegraphics[width=8cm]{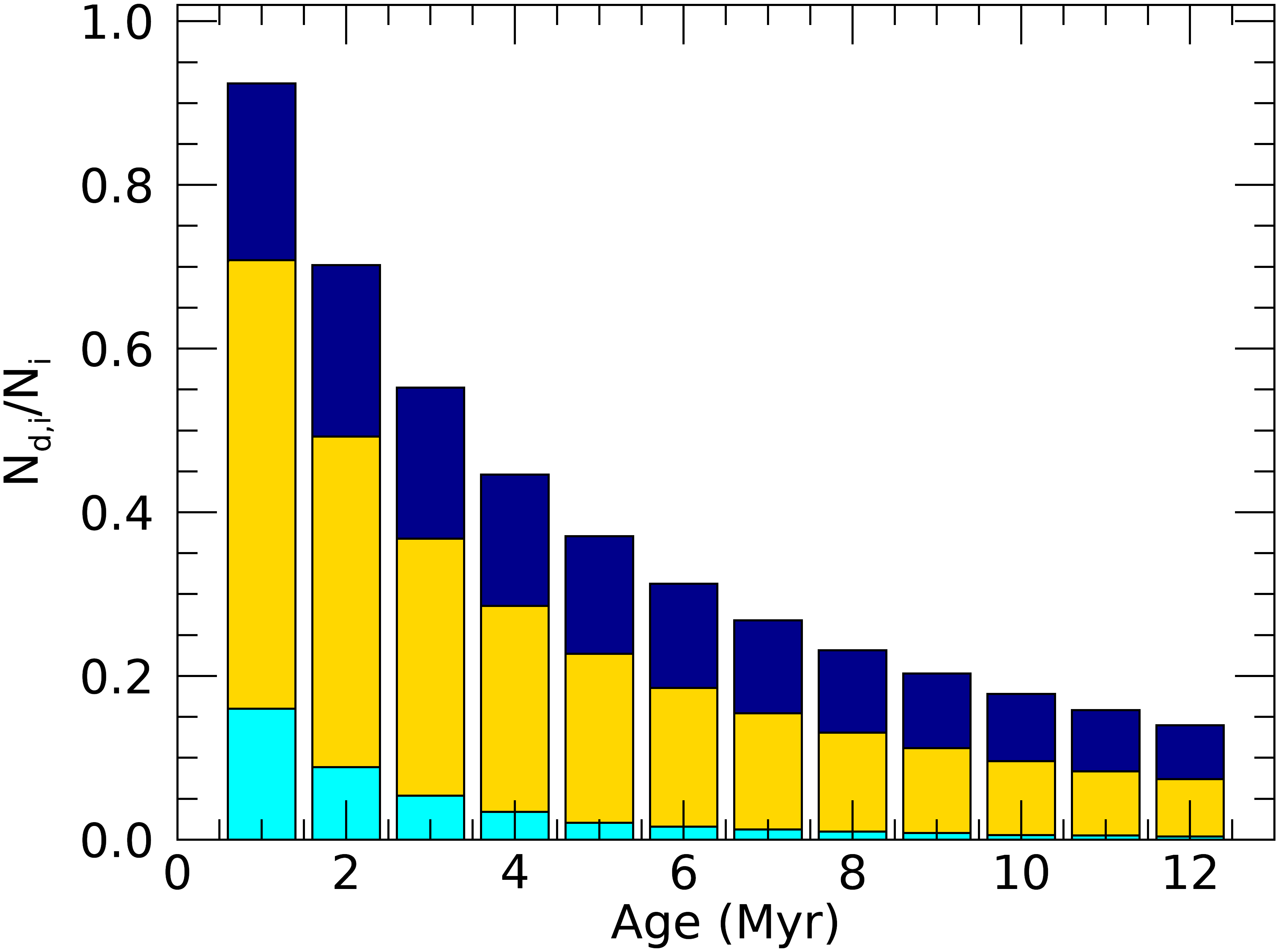}
  \caption{Top panel: Distribution of the parameter
  $\tau_\mathrm{th}$ for slow, median and fast rotators obtained
  for model M1. Bottom panel: Distribution of disk stars.
  Slow rotators are represented by dark blue bars; median rotators
  by yellow bars and fast rotators by light blue bars.}
  \label{tdisk_distPmed75}
\end{figure}

The equations are evolved from 1 to 550 Myr for models M1 and M2
and from 13 to 550 Myr for model M3. All the results are compared
to data of the ONC \citep{2014MNRAS.444.1157D, 2007ApJ...671..605C},
Cyg OB2 \citep{2017A&A...603A.106R}, Upper Sco \citep{2018AJ....155..196R},
h Per \citep{2013A&A...560A..13M}, the Pleiades
\citep{2016AJ....152..113R,2019A&A...628A..66L} and M37
\citep{2009ApJ...691..342H}. The selection criteria applied to the
observational data are shown in Table~\ref{tabcritobs}. The Cyg OB2
sample only contains stars with periods longer than 2 days because
they are free from contaminants \citep[for more details,
see][]{2017A&A...603A.106R}. For the Pleiades, we have cross matched
the samples from \citet{2019A&A...628A..66L} and
\citet{2016AJ....152..113R}.  In this work, we adopt solar metallicity
stellar evolutionary models \citep{1998A&A...337..403B}.  Most of
the samples analyzed here have solar metallicity. The ONC
\citep{1996ApJ...471..847P,2009A&A...501..973D,2011A&A...525A..35B}, U
Sco \citep{2009A&A...501..965V} and the Pleiades
\citep{2017PASJ...69....1T} present solar metallicity. Although Cyg
OB2 does not have determinations of abundance, \citet{2010ApJ...713..871W},
\citet{2013ApJ...773..135G} and \citet{2017A&A...603A.106R} all use
solar metallicity stellar models. \citet{2018A&A...620A..56B} find
no evidence of self-enrichment in Cyg OB2. There are some works
showing that the metallicity of h Per is sub solar \citep[Z = 0.01
-][]{2004MNRAS.349..547S,2011A&A...526A..76T}.  However,
\citet{1990A&A...232..431D} and \citet{1997ApJ...481L..47S} obtained
solar abundances for the same cluster.  For M37 (NGC2099),
\citet{2017MNRAS.470.4363C} obtained Z = + 0.08 while
\citet{2016A&A...585A.150N} obtained Z = + 0.02. There are several
uncertainties related to age determination particularly for young
clusters \citep[see, for example][]{2014prpl.conf..219S}. The ages
and corresponding references adopted in this work are shown in the
2$^\mathrm{nd}$ and 3$^\mathrm{rd}$ columns of Table~\ref{tabcritobs}.
For NGC2362, \citet{2008MNRAS.386..261M} suggest an age range between
4 and 5 Myr. We choose the age of 5 Myr because this is in better
agreement with our models. For Cyg OB2, \citet{2015MNRAS.449..741W}
observe an age spread consistent with an extended stellar formation
phase of 6 Myr with a peak between 4 - 5 Myr. We assume an age of
5 Myr for the cluster, as suggested by \citet{2010ApJ...713..871W}.
Among the clusters studied in this work, the age attributed to U
Sco exhibits the greatest variation, from 5 Myr
\citep{2008ApJ...688..377S,2015ApJ...808...23H} to 11 Myr
\citep{2012ApJ...746..154P}. \citet{2018AJ....155..196R} adopt 8
Myr for it but an 11 Myr old rotational period distribution returns
a better value in the KS tests performed in this work.

\subsection{Model M1} \label{M1}

In the top panel of Fig.~\ref{tdisk_distPmed75}, we show the
distribution of the parameter $\tau_\mathrm{th}$ obtained using
model M1. We define three classes of rotators, following
\citet{2013A&A...556A..36G}: fast rotators, with $P \le 1.4 \;
\mathrm{d}$; median rotators, with $1.4  \; \mathrm{d} \le P \le
10 \; \mathrm{d}$ and slow rotators, with $P > 10 \;\mathrm{d}$.
In this figure, there are two peaks for fast rotators, one at
$\boldmath{\tau_\mathrm{th}}$ equal to 0.67-0.97 Myr and the other
at 1.87-2.17 Myr both shorter than the peak for median rotators
which is between 2.17 - 2.47 Myr.  Most of slow rotators present
$\tau_\mathrm{th}$ values around $\sim 2.48 - 2.78$ Myr with a
maximum value around 3.4 Myr. As discussed in Section~\ref{model_setup},
higher $\tau_\mathrm{th}$ values imply lower
$\dot{\mathrm{M}}_\mathrm{acc,th}$. Because of this, most of the
fast rotators will have short living disks while median and slow
rotators will have longer living disks.  This can be seen at the
bottom panel of Fig.~\ref{tdisk_distPmed75} where the distribution
of disk stars is shown for slow, median and fast rotators. At 1
Myr, only 15\% of fast rotators have disks against more than 90\%
of the slow rotators. Only at 12 Myr the fraction of disk slow
rotators falls below the initial fraction of disk fast rotators.
At 12 Myr, all fast rotators are diskless while we can find around
13\% and 7\% of disk bearing stars among slow and median rotators,
respectively. In spite of that, after this age, all stars will be
considered diskless and their periods will be evolved following
equation~(\ref{wenv}) and equation~(\ref{wcore}).

Core-envelope coupling timescale values, $\tau_\mathrm{ce}$, are
shown in Table~\ref{initcond}. The values were taken from
\citet{2015A&A...577A..98G}. This timescale sets the rate of
angular momentum transfer from the core to the envelope. One can
note that it depends on mass and initial rotation rate. With these
values, 0.5 M$_\odot$ stars will take much longer to reach uniform
internal rotation than 1.0 M$_\odot$ stars.

\begin{figure*}
\centering
\includegraphics[width=\textwidth]{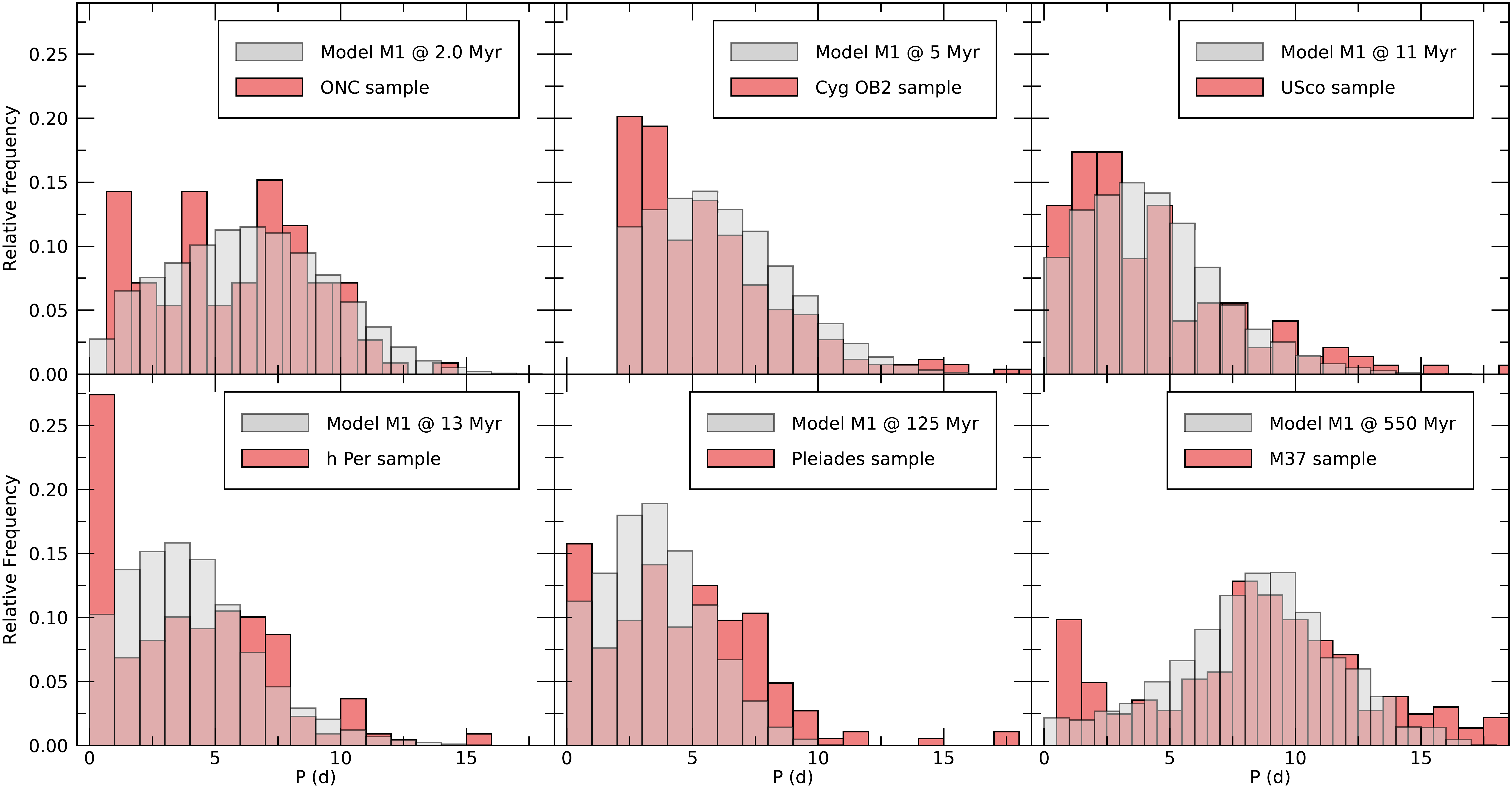}
\caption{Period distributions of model M1 at different ages
(grey bars) and the observational data (light coral bars).  Top
row: (left panel) model at 2.0 Myr superimposed to ONC period
distribution; (middle panel) model at 5 Myr superimposed to Cyg OB2
data and (right panel) model at 11 Myr over U Sco data. Bottom row:
(left panel) model at 13 Myr and h Per data; (middle panel) model
at 125 Myr and Pleiades data; (right panel) model at 550 Myr and
M37 data.}
\label{histPmed75}
\end{figure*}

\begin{figure*}
  \centering
  \includegraphics[width=\textwidth]{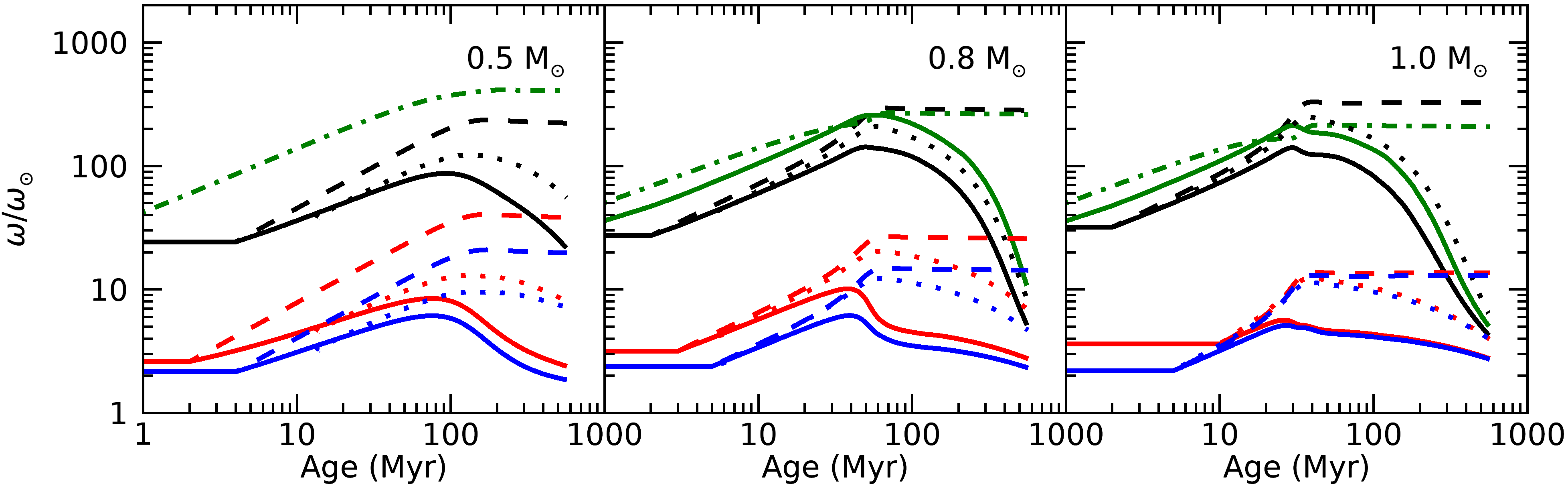}
  \caption{Angular velocity evolution as a function of age.
  Blue, red and black curves show the evolution of three randomly
  selected stars at each mass bin. The three chosen stars have disks
  at 1 Myr and are slow (blue), medium (red) and fast (black)
  rotators, respectively. Solid lines show the angular velocity of
  the convective envelope while dotted lines show the angular
  velocity of the radiative core. Dashed lines trace the evolution
  of angular velocity considering angular momentum conservation
  after disk loss following \PaperI's model M2. Green dash-dotted
  lines show the breakup limit and solid green lines show the most
  extreme rotators with 0.8 M$_\odot$ (middle panel) and 1.0 M$_\odot$
  (right panel). \label{wevol75b}}
\end{figure*}

For model M1, all the initial parameters have been chosen in order
to maximize the statistical compatibility with the ONC period
distributions. In the top left panel of Fig.~\ref{histPmed75} we
show the period distribution at 2.0 Myr obtained with this model
(grey bars) superimposed to observational ONC data (coral bars)
which is the result of the combination of the samples by
\citet{2007ApJ...671..605C} and \citet{2014MNRAS.444.1157D}. In the
same figure we present more evolved distributions. In the top row,
we show the distribution at 5 Myr compared to Cyg OB2 and the
distribution at 11 Myr compared to U Sco data. At the bottom row
we show the distributions at 13 Myr, 125 Myr and 550 Myr superimposed
to h Per, Pleiades and M37 samples, respectively. We see that the
theoretical distributions move to shorter periods increasing the
number of faster rotators from 2 Myr to 125 Myr. In fact, the peak
of the distribution moves from 5-8 days at 2 Myr to 3-4 days at 13
Myr and 125 Myr. However, the peak moves back to 8-10 days at 550
Myr which means that the number of slow rotators increases at later
ages, as can be seen in Fig.~\ref{wevol75b}.  Analyzing this figure,
we note that after the constant evolution at the disk phase, the
stars experience a strong spin up during the PMS. After that, the
loss of angular momentum due to the magnetized stellar wind causes
a strong braking and this moves the distribution to longer periods
(smaller angular velocities values) at 550 Myr. A very small fraction
of model M1 stars reach breakup velocities.  They represent at most
1\% of 1.0 M$_\odot$ and 0.3\% of 0.8 M$_\odot$ stars.  These
fractions decrease to 0.4\% of 1.0 M$_\odot$ and 0.2\% of 0.8
M$_\odot$ stars in model M2 (section 2.2.1) and to 0.4\% of the
stars in the mass range 0.6 - 1.0 M$_\odot$ in the last model
considered (section 2.2.3).  None of the stars of model M3 reach
or exceeds the critical limit.  There is no attempt to fix or to
remove these stars from the simulations because the number of
critical rotators is not high enough to alter the conclusions of
our study. In Figure 4, we have included three dash-dotted green
lines showing the breakup velocities and two solid green lines
showing the angular velocities of the 0.8 M$_\odot$ and 1.0 M$_\odot$
stars which present the most extreme rotation rates. The fastest
rotator is a 1.0 M$_\odot$ star that achieves at 28 Myr an angular
velocity that is 27\% higher than the expected breakup angular
velocity at this age followed by a 0.8 M$_\odot$ star that rotates
at a rate 12\% higher than the break up at 48 Myr.

\begin{table}
 \renewcommand{\arraystretch}{1.5}
  \centering
  \caption{K-S tests results comparing the simulated period
  distributions for all stars and the observational period
  distributions.}
   \begin{tabular}{|c|c|c|c|c|}
    \hline \hline
      Cluster & $D_\mathrm{crit}$ & $D_\mathrm{M1}$ & $D_\mathrm{M2}$
      & $D_\mathrm{M3}$ \\
      \hline
      ONC & 0.128 & 0.10 & 0.15 & - \\
      Cyg OB2 & 0.084 & 0.16 & 0.16 & - \\
      NGC 2362 & 0.14 & 0.11 & 0.12 & - \\
      U Sco & 0.11 & 0.11 & 0.21 & - \\
      h Per & 0.092 & 0.18 & 0.22 & 0.01 \\
      Pleiades & 0.096 & 0.22 & 0.14 & 0.17 \\
      M37 & 0.071 & 0.097 & 0.13 & 0.094 \\
    \hline
    \end{tabular}
    \label{KStests}  
\end{table}

\begin{figure*}
    \centering
    \includegraphics[width=\textwidth]{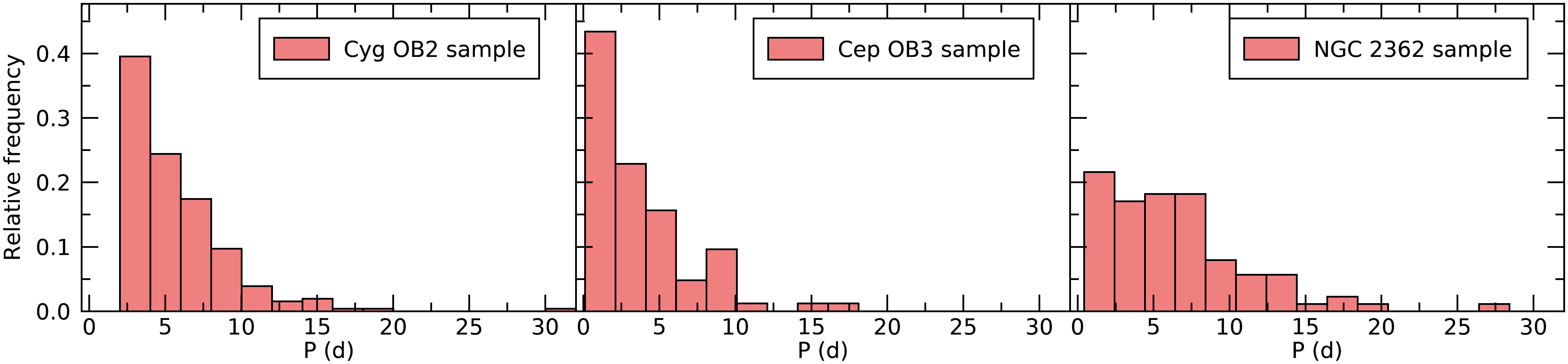}
    \caption{Period distributions of three 5 Myr old clusters:
    CygOB2 (left panel), CepOB3 (middle panel) and NGC 2362 (right
    panel).}
    \label{samples5myr}
\end{figure*}

In order to compare the samples, we apply a two-sample Kolmogorov-Smirnov
(K-S) test to the simulated and observational period distributions.
Through this test we calculate the maximal difference $D_{n,m}$
between two cumulative functions of size $n$ and $m$ and compare
it to a critical value $D_\mathrm{crit}$ that depends on the number
of objects on the samples and the significance level, taken here
to be equal to 0.05. If $D_{n,m} < D_\mathrm{crit}$, the simulated
and observational samples can be said to be derived from the same
population. In Table~\ref{KStests}, we show the K-S statistics
expressed by $D_\mathrm{crit}$ and by the $D$ values obtained for
models M1, M2 and M3, respectively.  From this table we note the
ONC sample and the simulated period distribution at 2 Myr for model
M1 are derived from the same population. We arrive at the same
conclusion for NGC 2362 and U Sco. For all the other clusters, the
comparison indicates that the agreement is not significant at the
0.05 level. Interestingly, while there is no similarity with Cyg
OB2 data, model M1 and NGC 2362 can be said to derive from the same
population, although both clusters have the same age. When we look
at the two distributions separately (Fig.~\ref{samples5myr}) we
note that Cyg OB2 has more fast rotating stars than NGC 2362 which
has relatively more slowly rotating ones. However, Cyg OB2 has two
stars rotating slowly than 30 days while the maximum period found
in NGC 2362 is around 27 days. \citet{2010MNRAS.403..545L} also
found that 5 Myr old Cep OB3b presents a different rotation pattern
when compared to NGC 2362. Its period distribution can also be seen
in Fig.~\ref{samples5myr}. The 3 clusters present different period
distributions, although Cyg OB2 and CepOB3 are more similar, with
more fast rotating stars than NGC 2362. The same is true for U SCo
and h Per, which are about the same age but exhibit quite different
rotational period distributions. As \citet{2010MNRAS.403..545L} and
\citet{2016ApJ...833..122C} pointed out there seems to exist
differences among clusters of same age due perhaps to environmental
causes. Recent papers
\citep{2021MNRAS.501.1782C,2021A&A...650A.157G,2021MNRAS.508.3710R} explore
different mechanisms that can influence the evolution of disks
depending on the cluster's environment. A shorter disk lifetime,
for example, can have an impact on the rotation pattern observed
in coeval clusters.

The comparisons between model M1 period distributions and the
observational samples older than U Sco show worse results.  At 13
Myr the agreement with the h Per sample is poor. The observational
sample is bimodal while the model is not. The number of the fastest
rotators ($\sim$ 1.0 day) and stars within the period range of 6-8
days is smaller than observed. On the other hand, there is an excess
of stars rotating at periods between 1-6 days. Indeed, the K-S
statistic for this case gives a maximal difference D$_{n,m}$ which
is two times greater than the critical one. The agreement with the
Pleiades is not better. The same deficiencies observed for the h
Per sample are seen here, but displaced to slightly shorter periods.
The comparison of model M1 at 550 Myr with M37 data is better but
we cannot say that the samples are drawn from the same population.
There is still a lack of fast rotators and of stars with periods
longer than 14 days and an excess of stars with periods between 4
and 11 days.

We can note the effects the introduction of the mechanisms of
angular momentum variation produced in our models looking at Figure
\ref{wevol75b}, where dashed lines trace the evolution of the angular
velocity of randomly chosen stars considering angular momentum
conservation similarly to \PaperI's model M2. The curves show no
variation after an initial steep increase, just after the disk loss,
when the star arrives at the ZAMS. Statistical comparisons of period
distributions obtained under this J constant condition with the
observational samples of stellar clusters younger than 13 Myr give
worse results than those we have obtained with model M1: only at
2.0 Myr model and observational distributions can be said to be 
derived from the same population. Even before the arrival at the 
main sequence the angular momentum internal re-distribution and
loss play an important role in determining the rotational behavior
of a star.

\begin{figure*}
\centering
\includegraphics[width=\textwidth]{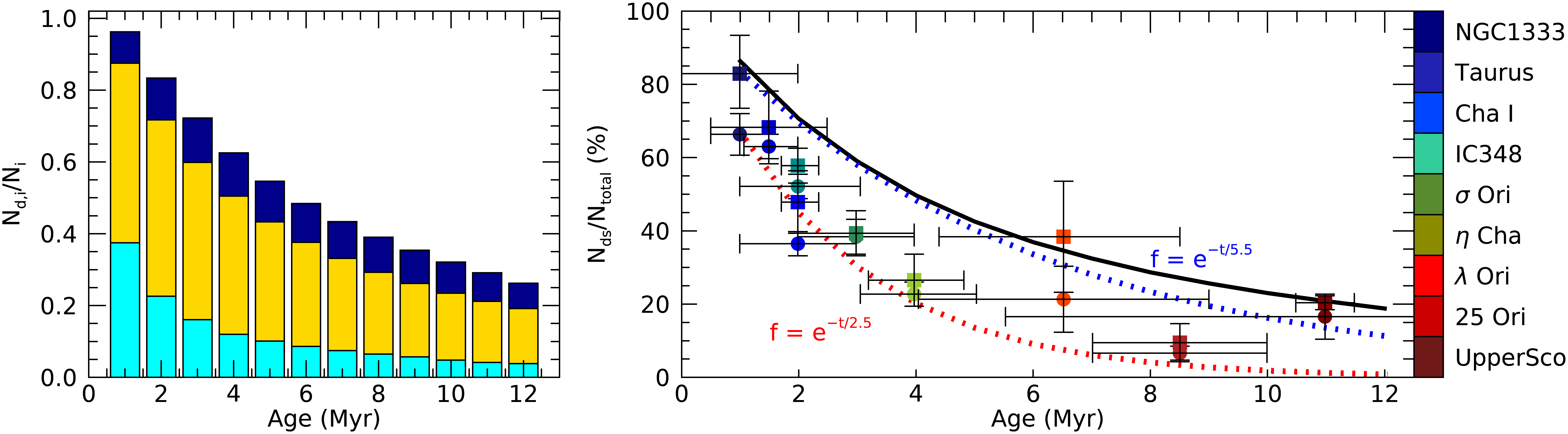}
\caption{Left panel: Distribution of disk stars for slow (dark
blue), median (yellow) and fast (light blue) rotators for model
M2. Right panel: disk fraction as a function of age for the same
model; lines and symbols are the same of those seen at the right
panel of Fig.~\ref{tth75}.  Results obtained using $\overline{P} =
2.0$ d.}
\label{disklifPmed2}
\end{figure*}

\begin{figure*}
\centering
\includegraphics[width=\textwidth]{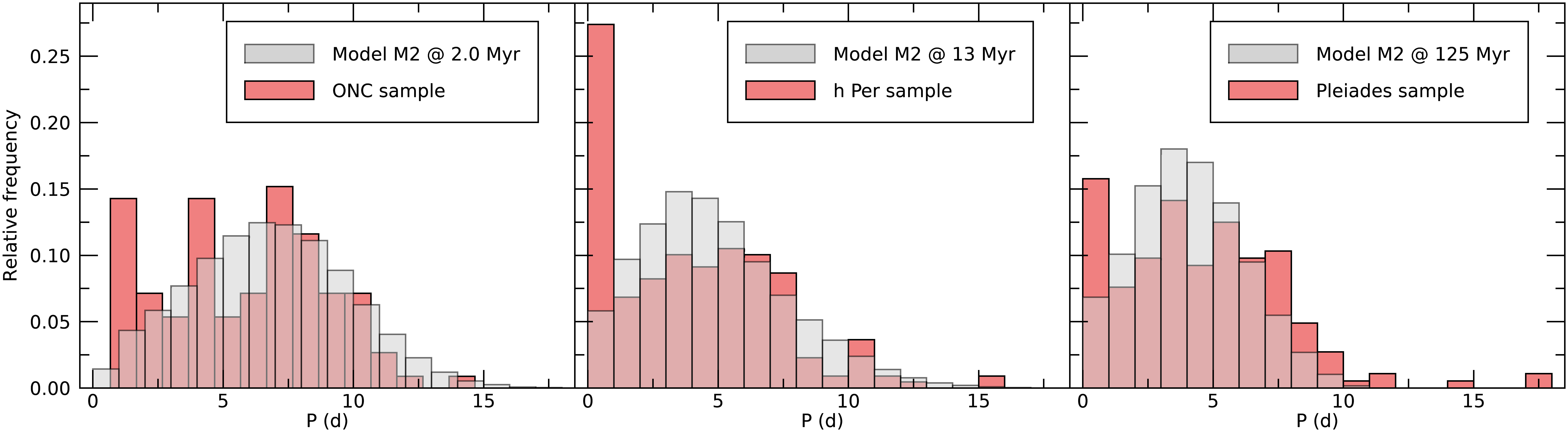}
\caption{Same as Fig.~\ref{histPmed75} but for model M2.}
\label{histPmed2}
\end{figure*}

\subsection{Exploring other parameters - models M2, M3 and other
initial mass distributions} \label{difparam}

In order to try to understand the impact of some key parameters on
the simulations, we will explore a different set of values. We will
change the parameter $\overline{P}$ and the initial conditions.
Also, we will consider initial mass distributions based on the
samples studied.

\subsubsection{Model M2} \label{modM2}

Since model M1 was not able to explain the evolution of the rotational
period distributions from the PMS to MS phases we decided to explore
other parameters. In model M2, we change the value of $\overline{P}$ to
2.0 days.  This change will increase the maximum $\tau_\mathrm{th}$
value from 3.4 days to 7 days causing a reduction of the
mass accretion rate threshold values, increasing the disk fractions
of the rotators (left panel of Fig.~\ref{disklifPmed2}) and
the total disk fraction that now falls off more slowly than
previously seen in model M1 (Fig.~\ref{disklifPmed2}, right panel).

Stars with longer living disks will maintain their initial rotational
rates for a longer time \citep{2015A&A...577A..98G}.  This will
increase the number of slower rotators which is what effectively
happened as can be seen in Fig.~\ref{histPmed2} where period
distributions at 2.0 Myr, 13 Myr and 125 Myr are shown in comparison
with data from the ONC, h Per and the Pleiades. When we compare
this figure to Fig.~\ref{histPmed75}, at 2 Myr, the number of stars
with $P \le 5$ days has decreased while it has increased for 5 d
$< P \le 12$ d. At 13 Myr and 125 Myr, one can observe the same
behavior but at different intervals: at both ages, the number of
stars decreases in comparison with the standard model until P = 4
d; at 13 Myr, it increases over the range from 6 d to 15 d; at 125
Myr over the range from 5.0 d to 11 d. The 4$^\mathrm{th}$ column of
Table~\ref{KStests} indicates that the maximal difference increased
for almost all clusters which means that the agreement among the
M2 simulated period distributions and the observational clusters
is now poorer compared to that of model M1, except for Cyg OB2 and
the Pleiades and also for NGC 2362, because the D$_\mathrm{n,m}$
increase was very small in this case.  For Cyg OB2, the maximal
difference did not change but for the Pleiades it decreased
significantly  but it is still greater than the critical value.

\subsubsection{Different $\tau_\mathrm{ce}$ values} \label{diftau}

\begin{figure}
\centering
\includegraphics[width=0.45\textwidth]{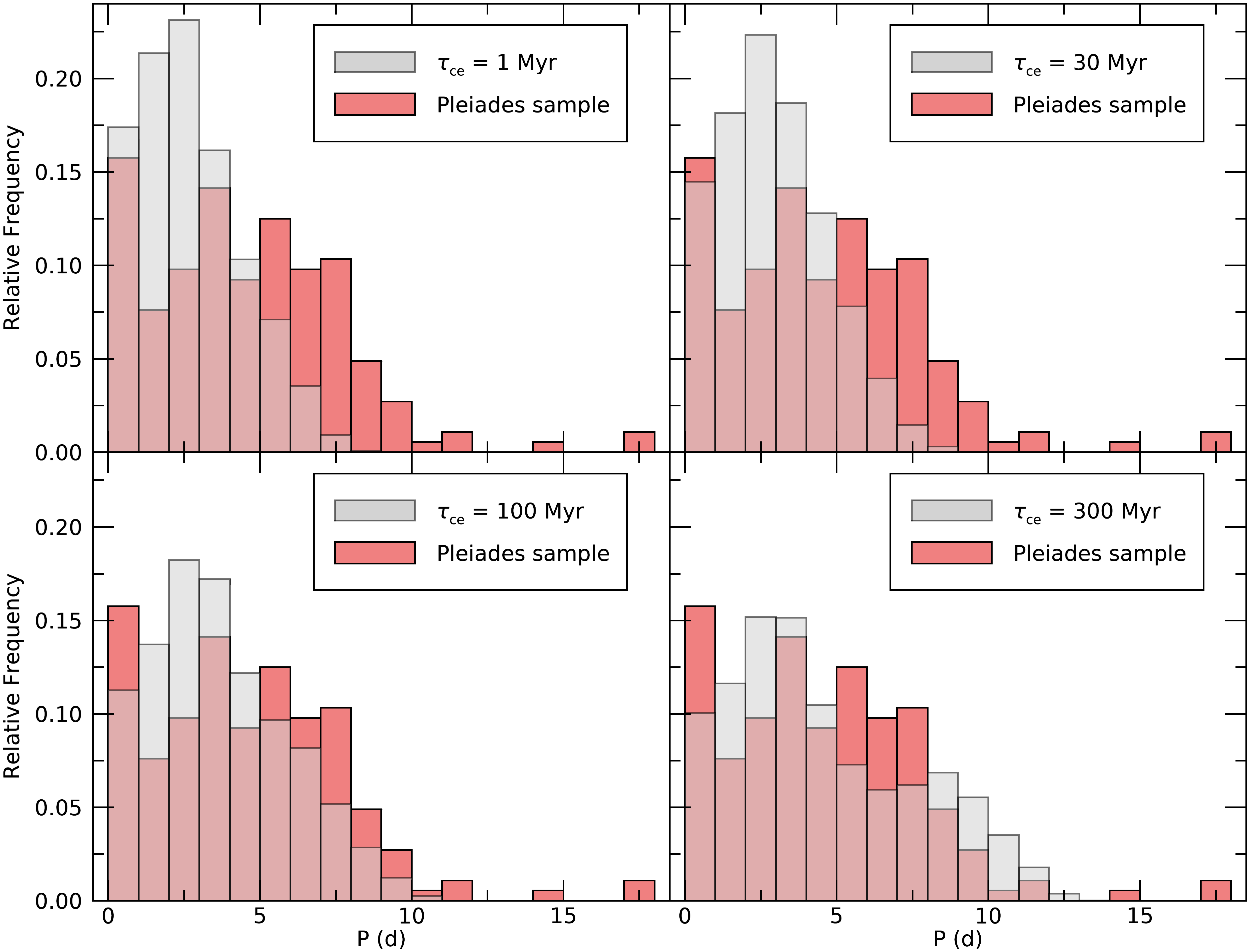}
\caption{Simulated period distributions at 125 Myr obtained
with $\tau_\mathrm{ce}$ equal to 1 Myr (top left panel), 30 Myr
(top right panel), 100 Myr (bottom left panel) and 300 Myr (bottom
right panel) superimposed to observational data of the Pleiades.}
\label{figdiftauPlei}
\end{figure}

We tried different $\tau_\mathrm{ce}$ values in order to analyze
the impact this parameter has in our simulations. We run 4 simulations
using model M1 parameters but with $\tau_\mathrm{ce}$ equal to 1.0,
30, 100 and 300 Myr for all rotators and one simulation with
$\tau_\mathrm{ce}$ equal to 300 Myr for slow and medium rotators
and equal to 1 Myr for the fast rotators.  Our results show that
the period distributions from 2 to 13 Myr are barely affected by
the assumed $\tau_\mathrm{ce}$ values. The K-S tests that compare
the synthetic period distributions with the observed ones return
practically the same D values at this age range (Table~\ref{KSteststau}).
In Figure \ref{figdiftauPlei} we show period distributions obtained
at 125 Myr for the different $\tau_\mathrm{ce}$ values. One can see
that at this age, low $\tau_\mathrm{ce}$ values (1 Myr and 30 Myr)
increase the number of short period rotators (P $< 4$ d) and decrease
the number of slowing rotating stars. An intermediate value of
$\tau_\mathrm{ce} = 100$ Myr decreases the number of stars with 3
d $< P \le 6$ d and increases the number of longer period rotators.
Finally, with a value of $\tau_\mathrm{ce} = 300$ Myr, the number
of stars with P $> 4$ d increases. Shortly, higher $\tau_\mathrm{ce}$
values increases the number of slow rotators at the age of 125 Myr.
The lowest $\tau_\mathrm{ce}$ value (equal to 1 Myr) does not produce
good KS values neither at 125 Myr nor at 550 Myr. Compared to those
seen in Table~\ref{KStests}, the results improve at 125 Myr but are
worse at 550 Myr. When using $\tau_\mathrm{ce}$ = 100 Myr the period
distribution at 550 Myr has a peak around 8 days and more intermediate
rotators than observed in M37 period distribution. This increases
the difference between the observed and the simulated cumulative
distributions in comparison to what is obtained using the values
seen in Table~\ref{initcond}.  On the other hand the run with
$\tau_\mathrm{ce}$ = 300 Myr for slow and medium rotators and
$\tau_\mathrm{ce}$ = 1 Myr for fast rotators gives the same D values
obtained from model M1 except for the Pleiades. The period distribution
at 125 Myr is much more sensitive to the value of $\tau_\mathrm{ce}$
than the period distributions at older ages but in general, the
higher the $\tau_\mathrm{ce}$ the better the results. We can conclude
that the $\tau_\mathrm{ce}$ values (Table~\ref{initcond}) proposed
by \citet{2015A&A...577A..98G} are adequate to reproduce the cluster's
period distributions we use for comparison in this work.  We can
also conclude, as already stated by
\citet{2013A&A...556A..36G,2015A&A...577A..98G}, that solid body
rotation promotes fast rotation at the ZAMS.

\begin{table}
 \renewcommand{\arraystretch}{1.5}
  \centering
  \caption{K-S tests results comparing the simulated period
  distributions and the observational period distributions using
  different $\tau_\mathrm{ce}$ values.}
   \begin{tabular}{|c|c|c|c|c|c|}
    \hline \hline
      Cluster & \multicolumn{5}{|c|}{$\tau_\mathrm{ce}$ values (Myr)} \\
              & 1 & 30 & 100 & 300 & 300, 300, 1 \\
      \hline
      ONC & 0.08 & 0.09 & 0.09 & 0.09 & 0.08 \\
      Cyg OB2 & 0.15 & 0.15 & 0.15 & 0.15 & 0.15 \\
      NGC 2362 & 0.11 & 0.11 & 0.11 & 0.11 & 0.11 \\
      U Sco & 0.10 & 0.11 & 0.10 & 0.11 & 0.10 \\
      h Per & 0.18 & 0.18 & 0.18 & 0.18 0.18 \\
      Pleiades & 0.33 & 0.30 & 0.17 & 0.08 & 0.08 \\
      M37 & 0.33 & 0.09 & 0.15 & 0.097 & 0.10 \\
    \hline
    \end{tabular}
    \label{KSteststau}
\end{table}

\subsubsection{Model M3} \label{modM3}

Examining Table~\ref{KStests}, we note that model M1 is successful
in explaining three out of four clusters younger than h Per but it
cannot reproduce the older clusters taken for comparison in this
work. We can wonder if the problem is related to our initial
conditions, to the assumed disk locking mechanism, to the wind
prescription or to the core-envelope angular momentum exchange. In
order to rule out the first two uncertainties, we decided to run
model M3 similarly to what was done by
\citet{2016ApJ...833..122C}. This model begins at 13 Myr and
its initial period distribution is taken from the observational
period distribution of h Per \citep{2013A&A...560A..13M} but
otherwise it is similar to model M1, using the same wind prescription
and the same core-envelope coupling timescale values.

\begin{figure*}
\centering
\includegraphics[width=\textwidth]{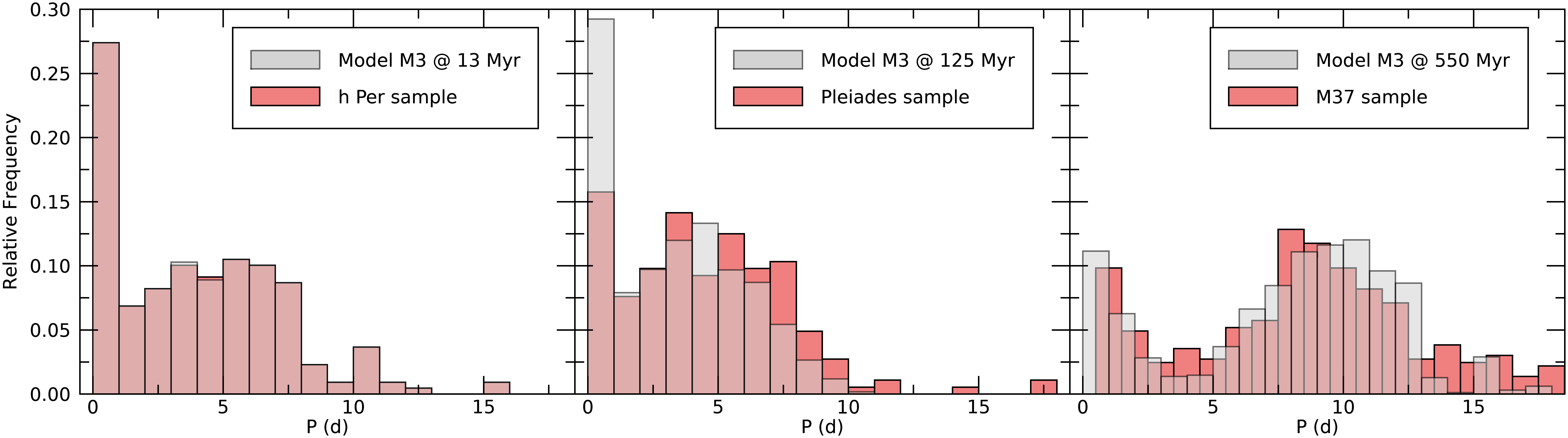}
\caption{Simulated period distributions at 13 Myr, 125 Myr and 550
Myr obtained with model M3 superimposed to observational data of h
Per, the Pleiades and M37.}
\label{histM3}
\end{figure*}

In Fig.~\ref{histM3} we show the period distributions obtained
with model M3 at 13 Myr, 125 Myr and 550 Myr compared to observational
distributions of h Per, the Pleiades and M37.

At 13 Myr, the simulated and the observational distributions are
almost perfectly superimposed. The maximal difference between the
two cumulative distributions seen at the 5$^{th}$ column at
Table~\ref{KStests} is 0.01, much smaller than the critical one.
Obviously, the two samples are derived from the same population as
imposed by the initial conditions of this model. Even so, the
Pleiades and M37 period distributions can not be reproduced by this
model according to the values obtained from the KS tests. Visually
however the simulated period distributions are much similar to the
observed ones. The number of fast rotators increased a lot compared
to what was obtained with the other models. At the age of the
Pleiades, there is an excess of simulated stars with periods shorter
than 1.0 day and between 2.0 and 5.0 days compared to the observational
sample.  On the other hand, there is a lack of simulated stars at
longer periods. The simulated period distribution at the age of 550
Myr is very similar to the observed one.

We can conclude that the mechanisms (the wind mass loss prescription
and/or the core-envelope coupling parameterization) and the parameters
applied to this model produced a very good visual match of the
distributions but not quantitatively enough to fully explain the
observed period distributions. For MS clusters as old as M37 (550
Myr old) initial conditions seems to play a secondary role in the
evolution of the angular momentum.

\subsubsection{Other mass distributions} \label{MF}

\begin{figure}
\centering
\includegraphics[width=0.45\textwidth]{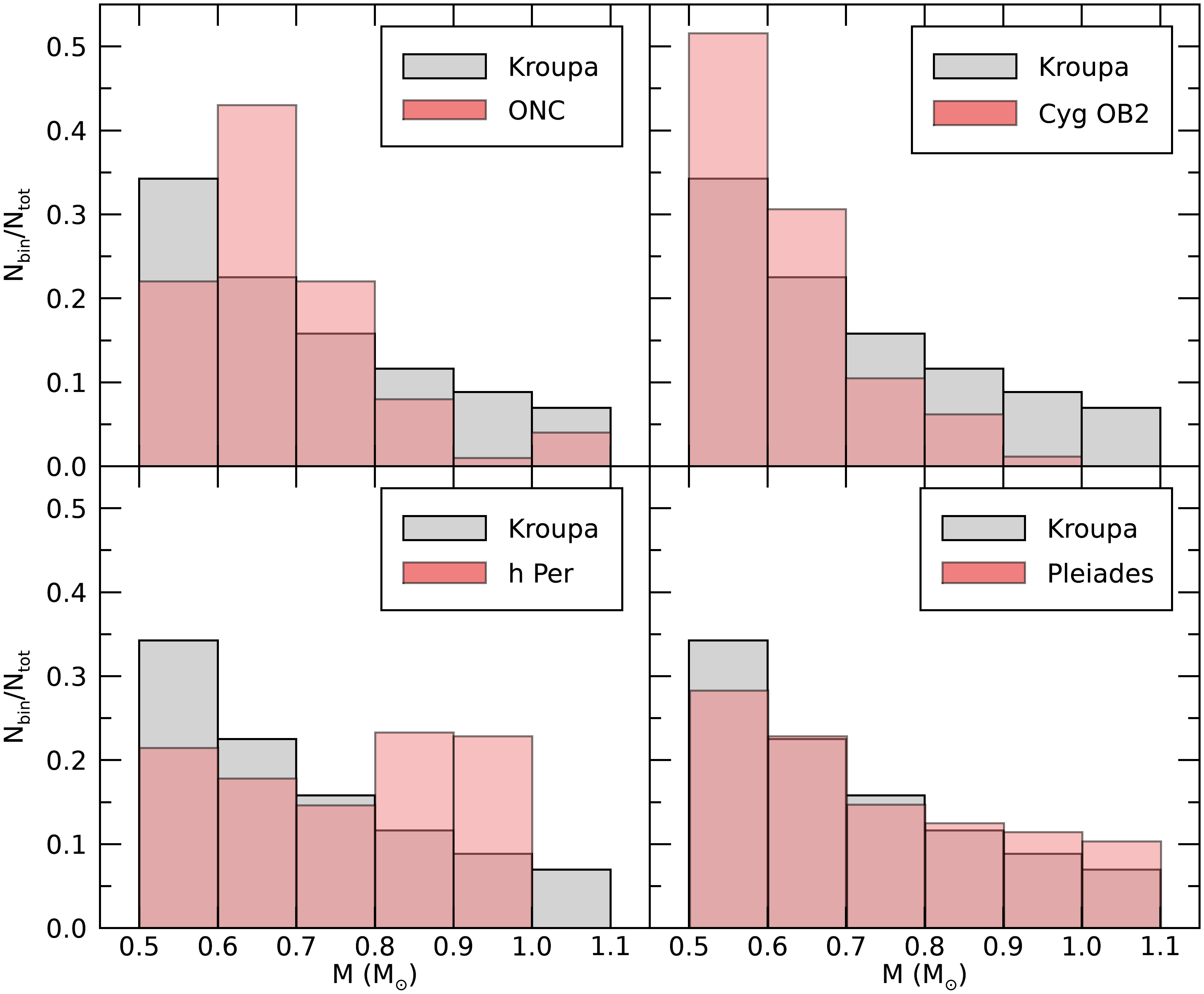}
\caption{Comparison of mass distributions of the ONC, Cyg OB2, h Per
and the Pleiades to Kroupa's IMF.}
\label{figMdist}
\end{figure}

Young open clusters appear to have similar intrinsic mass functions
\citep{2010ARA&A..48..339B,2021arXiv210108804D}. However, the
observed samples do not evenly sample the intrinsic IMF. As a result,
the mass distribution of the samples we use here does not exactly
follow the underlying IMF. In order to take this observational bias
into account, we computed models similar to model M1 with a finer
mass grid reproducing the mass distributions of the observed samples
used in this work. Now the mass range goes from 0.5 to 1.1 M$_\odot$
in intervals of 0.1 M$_\odot$ values.  The percentage value at each
mass bin is shown in Table~\ref{tabMdist} and can be seen in
Fig.~\ref{figMdist}.

\begin{table}
\renewcommand{\arraystretch}{1.0}
 \begin{center}
  \caption{Kroupa's IMF and mass distributions of the selected young clusters.}
  \resizebox{0.5\textwidth}{!}{ %
   \begin{tabular}{|c|c|c|c|c|c|c|c|}
    \hline \hline
      MF & 0.5 M$_\odot$ & 0.6 M$_\odot$ & 0.7 M$_\odot$ & 0.8 M$_\odot$ &
         0.9 M$_\odot$ & 1.0 M$_\odot$  & Reference \\
    \hline
         & \multicolumn{7}{|c|}{\%} \\
    \hline
      Kroupa & 34 & 22 & 16 & 12 & 9 & 7 & 1 \\
      ONC & 22 & 43 & 22 & 8 & 1 & 4 & 2, 3 \\
      Cyg OB2 & 51 & 28 & 13 & 6 & 1 & 1 & 4 \\
      h Per & 17 & 14 & 12 & 18 & 18 & 20 & 5 \\
      Pleiades & 28 & 23 & 14 & 13 & 12 & 10 & 6, 7 \\
    \hline
    \end{tabular}}
    \end{center}
    \label{tabMdist}
    {\footnotesize (1) \citet{2013pss5.book..115K}, (2) \citet{2007ApJ...671..605C},
(3) \citet{2014MNRAS.444.1157D}, (4) Roquette et al. (2017), (5) Moraux et al. (2013),
(6) Rebull et al. (2016), (7) Lodieu et al. (2019).}
\end{table}

The values at each mass bin for the Kroupa mass function were
calculated using \citet{2013pss5.book..115K} equations for the
canonical IMF. The percentage of stars that appears in
Table~~\ref{tabMdist} was then calculated considering the total
number of stars of the simulation. To calculate the other mass
distributions we take the observational samples used in this work
for which we have stellar mass values (a combined sample of
\citet{2007ApJ...671..605C} and \citet{2014MNRAS.444.1157D} for the
ONC; Cyg OB2 from \citet{2017A&A...603A.106R}, h Per from
\citet{2013A&A...560A..13M} and a combined sample of
\citet{2016AJ....152..113R} and \citet{2019A&A...628A..66L} for the
Pleiades) and estimate the percentage of stars in the mass bins
from the total number of stars in each sample.  We then use these
percentages to calculate the amount of objects in the mass bins for
each one of the mass distributions shown in Table \ref{KSmassdist}.
Since the methods or models used by the different authors to assign
mass to the observational samples are not the same the sample used
here is not homogeneous.

\begin{figure}
\centering
\includegraphics[width=0.5\textwidth]{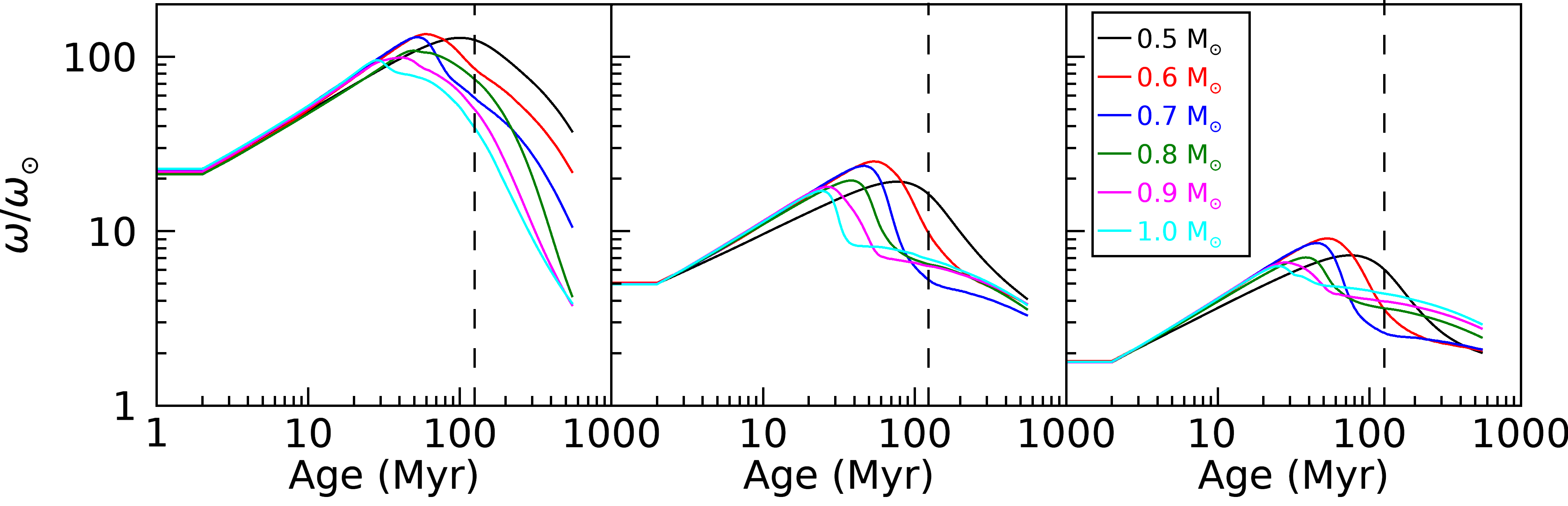}
\caption{Angular velocity evolution as a function of age considering
stars with masses between 0.5 and 1.0 M$_\odot$ and with a disk
lifetime equal to 1 Myr. In the left panel, all stars are fast
rotators (see text for a definition) and have approximately the
same initial rotational period. In the middle and right panels we
have, respectively, median and slow rotators. The vertical dashed
line mark the age of 125 Myr.}
\label{wevol6mass}
\end{figure}

When we analyze the angular velocity evolution for stars with
different masses but with the same initial rotation periods and
disk lifetimes at the age of 125 Myr, we note that their values do
vary with mass (Figure \ref{wevol6mass}). Considering stars in all
the three initial rotating intervals (fast, median and slow rotators),
the fastest rotating stars at 125 Myr are those of 0.5 M$_\odot$.
However there is no monotonic relation between mass and rotation.
Analyzing the fast rotators set, after the 0.5 M$_\odot$ star,
follows stars of 0.6, 0.8, 0.7, 0.9 and 1.0 M$_\odot$ in order of
decreasing angular velocity.  In the group of median rotators, the
sequence of decreasing angular velocity is again of stars of 0.5
and 0.6 M$_\odot$ and then stars of 1.0, 0.8 and 0.9 and 0.7
M$_\odot$.  For the slow rotators, it is 0.5, 1.0, 0.9, 0.8 and 0.6
and 0.7 M$_\odot$. Clearly, 0.5 M$_\odot$ stars stay longer at the
PMS phase, spinning up due to radius contraction. At 125 Myr, they
still present a high spin rate.

\begin{figure*}
\centering
\includegraphics[width=\textwidth]{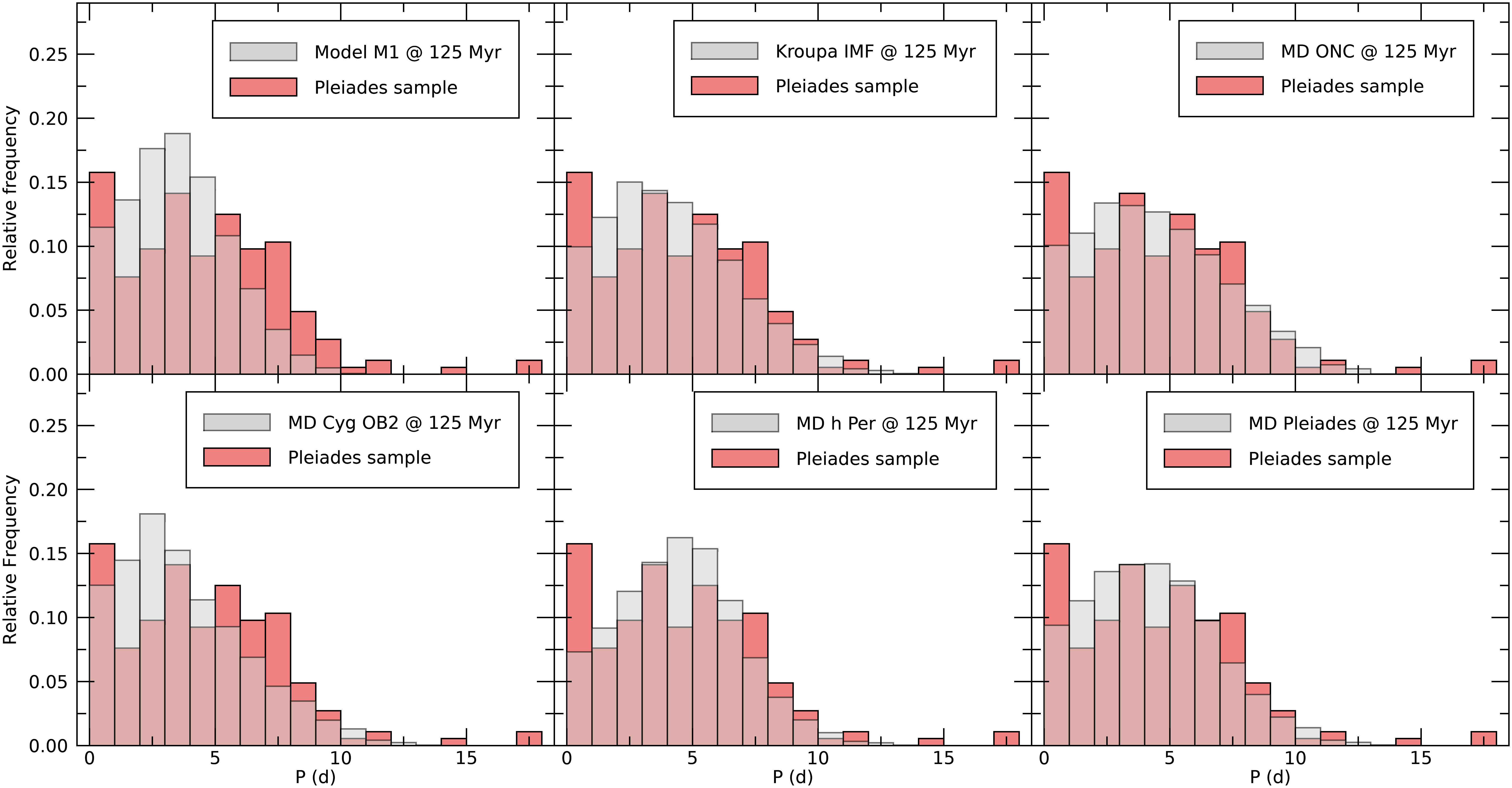}
\caption{Rotational period distributions at 125 Myr obtained
with model M1 and the mass distributions show in Table \ref{tabMdist}
and in Figure \ref{figMdist} all of them superimposed on Pleiades
observational period distribution. Top panels from left to right:
model M1 and models using Kroupa IMF with 7 mass bins and the ONC
MD. Bottom panels, from left to right: models using Cyg OB2, h Per
and Pleiades mass distributions.}
\label{hist_Plei_ONC_Mdist}
\end{figure*}

Therefore, mass distributions (MD) that have a high fraction of low
mass (0.5 and 0.6 M$_\odot$) stars will present a higher number of
stars with rotational periods below 1.0 day (the first bin of the
rotational period distributions - Figure \ref{hist_Plei_ONC_Mdist})
although stars at this mass range also produce the fastest rotators
of the other rotational intervals (medium and slow).  This is the
case of the results obtained with the Cyg OB2 MD and the Kroupa IMF
in model M1 (Figure \ref{figMdist} and the left panels of Figure
\ref{hist_Plei_ONC_Mdist}).  On the opposite side, the period
distribution obtained with a model that uses the h Per MD that has
a deficit of 0.5 M$_\odot$ star shows the lowest number of stars
with periods shorter than 1.0 d.  The results obtained using the
Pleiades MD that is very similar to Kroupa IMF with 7 mass bins
produces a different period distribution, however, due to the lower
number of 0.5 M$_\odot$ stars and a slightly higher fraction of 0.8
- 1.1 M$_\odot$ stars.  The objects that are initially classified
as medium rotators, at 125 Myr show periods between $\sim$ 2 and 5
days and the initially slow rotators populate the longer period
bins.  The period distribution obtained with the ONC MD is not so
visually different but produces statistically different values from
that obtained with the Kroupa IMF (Table \ref{KSmassdist}).  The
ONC MD has less 0.5 M$_\odot$ stars but an excess of 0.6 and 0.7
M$_\odot$ stars compared to the Kroupa IMF (Figure \ref{figMdist}).

In Table~\ref{KSmassdist} the K-S statistics of the models with
different MD is shown. One thing to be noticed is that the K-S
distances obtained using Kroupa's IMF (2$^\mathrm{nd}$ column) are
slightly different from model M1's except for the Pleiades for which
they are much different and smaller. Although none of the mass
distributions are similar to Kroupa's IMF, the critical distances
obtained from K-S tests are alike among the pre main-sequence
clusters (from ONC to h Per) and comparable to the values obtained
from model M1. For the Pleiades they are much smaller and different
for the various mass distributions.  Now, statistically there is
an agreement between the models at 125 Myr and the observed Pleiades
period distribution (but not for the model that uses Cyg OB2 mass
distribution). For M37, the K-S distance values are similar to those
obtained with model M1 but they are more sensitive to the different
mass distributions compared to the PMS clusters.

\begin{table}
 \renewcommand{\arraystretch}{1.5}
  \centering
  \caption{K-S tests results comparing the simulated period
  distributions for all stars and the observational period
  distributions using different IMF.}
   \begin{tabular}{|c|c|c|c|c|c|c|}
    \hline \hline
      MF & Kroupa$^\ast$ & ONC & Cyg OB2 & h Per & Pleiades & \\
    \hline
      Cluster & \multicolumn{5}{|c|}{D$_\mathrm{KS}$} & $D_\mathrm{crit}$ \\
      \hline
      ONC & 0.08 & 0.08 & 0.08 & 0.08 & 0.08 & 0.128 \\
      Cyg OB2 & 0.15 & 0.14 & 0.15 & 0.17 & 0.16 & 0.084 \\
      NGC 2362 & 0.13 & 0.14 & 0.13 & 0.14 & 0.10 & 0.14 \\
      U Sco & 0.095 & 0.097 & 0.095 & 0.10 & 0.10 & 0.11 \\
      h Per & 0.17 & 0.17 & 0.18 & 0.16 & 0.17 & 0.092 \\
      Pleiades & 0.097 & 0.07 & 0.16 & 0.10 & 0.08 & 0.096 \\
      M37 & 0.099 & 0.094 & 0.086 & 0.14 & 0.10 & 0.071 \\
    \hline
    \end{tabular}
    {\small $^\ast$ The difference between the K-S statistical for
    the model using Kroupa's IMF and that seen in Table~\ref{KStests}
    is due to the difference on the models mass range.}
    \label{KSmassdist}
\end{table}

\begin{figure*}
\centering
\includegraphics[width=\textwidth]{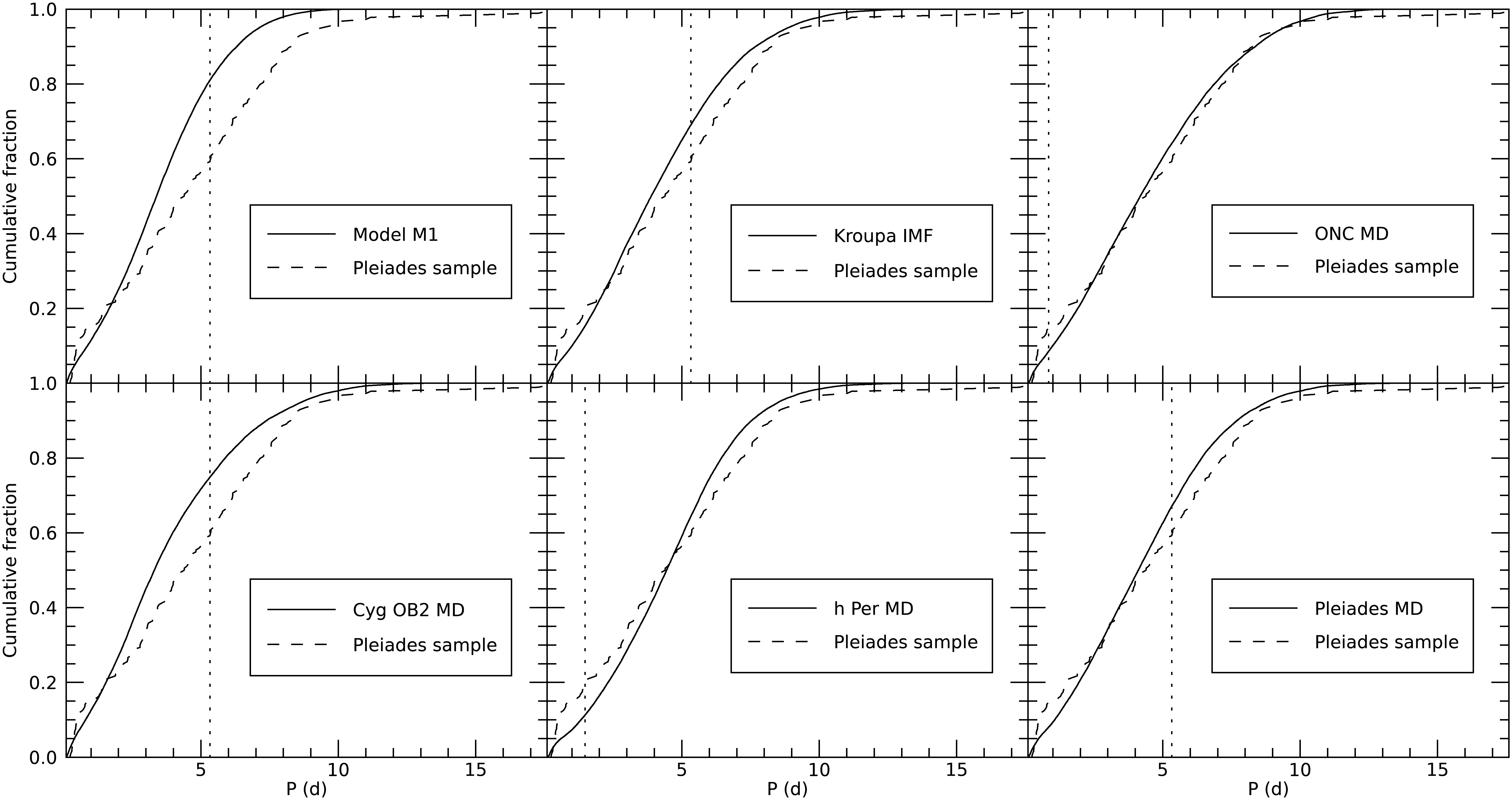}
\caption{Cumulative period distributions at 125 Myr obtained
with model M1 and the mass distributions show in Table \ref{tabMdist}
and in Figure \ref{figMdist} (solid lines) all of them superimposed
on Pleiades observational cumulative period distribution (dashed
lines). Top panels from left to right: model M1 and models using
Kroupa IMF with 7 mass bins and the ONC MD. Bottom panels, from
left to right: models using Cyg OB2, h Per and Pleiades mass
distributions. Also shown is a vertical line marking the corresponding
period where the maximal difference between of simulated and
observational samples occurs (dotted lines).}
\label{CPD}
\end{figure*}

In Figure \ref{CPD}, we show the cumulative distributions (CD)
relative to the period distributions seen in Figure
\ref{hist_Plei_ONC_Mdist}, with the point of maximal distance between
the observational Pleiades period distribution and the models marked
by a vertical dotted line. The conclusions stated previously appear
more clearly after an inspection of Figure \ref{CPD}. We see that,
despite the fact that in model M1 and in this section we present
results that uses the same IMF of Kroupa, the absence of some of
the mass bins makes the cumulative distributions different
(1\textsuperscript{st} and 2\textsuperscript{nd} top panels of
Figure \ref{CPD}). We also note that the CD obtained using model
M1 and Cyg OB 2 MD are very similar, with the maximal difference
between the observed and the model theoretical cumulative distributions
occurring at the same period. However, in model M1 there is an
excess of objects with periods longer than 3 days and this causes
the saturation of CD earlier than in the Cyg OB2 CD. There is a
deficit of objects with short rotational periods and an excess of
longer period objects when we examine the CD obtained through the
model with the h Per MD.  Also, we can clearly see why the model
with the ONC MD produces better statistical results than those using
Kroupa IMF and Pleiades MD at 125 Myr since the simulated and the
observational cumulative distributions fit very well.

\section{Conclusions} \label{conclusions}

In this work, we analyzed the rotational evolution of solar mass
stars from 1 Myr to 550 Myr taking into account disk locking,
core-envelope decoupling and angular momentum loss via a magnetized
wind.

Our standard model, model M1, reproduces well the disk fraction as
a function of age and the rotational period distributions of PMS
clusters younger than 13 Myr old.

We test the role the core-envelope coupling timescale has in our
simulations. At the pre-main sequence and ZAMS this parameter does
not seem to influence greatly the period distributions. It is more
important at the MS and it is necessary to have a weak coupling
(high $\tau_\mathrm{ce}$ values) for slow and medium rotators in
order to reproduce the observations.

We also run simulations assuming different values for the disk
lifetimes and for the initial period distributions. While these
models provide synthetic rotational period distributions resembling
those of observed clusters, they do not fulfill the quantitative
similarity criteria set by the K-S test. Running a finer grid model
and replacing the Kroupa's initial mass function by the empirical
mass functions of the samples studied here did not produce any
significant improvement compared to model M1 except for the Pleiades
that now are reproduced by our model.

It seems that there are some intrinsic differences among the clusters
related to their initial conditions. For example, enhanced disk
dissipation in massive clusters, multiplicity fraction, planet
formation in disks and the role magnetic fields play in the first
stages of the stellar formation may change the initial rotation
pattern of the population of different clusters. This could explain
the different rotation patterns observed in the clusters of about
the same age Cyg OB2 and NGC 2362 or U Sco and h Per.  These aspects
are interesting and will be analyzed in future works as well as the
impact of the use of different stellar models that take into account
accretion and rotation and the lithium-rotation connection problem.

We can conclude that in a global perspective the secular evolution
of rotation described by the models presented here grasp the main
trends of the spin evolution of low-mass stellar populations.
However specific clusters may keep signatures of their initial
conditions up to at least the ZAMS as it seems to be case for the
pre-main sequence clusters analyzed in this work, from the ONC to
h Per. Eventually, a physical description of the star-disk interaction
process that prevents PMS stars from spinning-up during the disk
locking phase remains to be provided.

\section*{Acknowledgments}

M.J.V. would like to thanks the Brazilian agency CAPES and the
PROPP-UESC (project 073.6766.2019.0010667-91) for partial funding.
J. B. acknowledges the support of the European Research Council
(ERC) under the European Union’s Horizon 2020 research and innovation
program (grant agreement No 742095 ; SPIDI : Star-Planets-Inner
DiskInteractions; http://www.spidi-eu.org). The authors thank an
anonymous referee for very useful suggestions that helped to improve
the quality of the paper.

\section*{Data Availability}

The data underlying this article will be shared on reasonable request
to the corresponding author.

\bibliographystyle{mnras} 
\bibliography{ref}

\begin{thebibliography}{}
\makeatletter
\relax
\def\mn@urlcharsother{\let\do\@makeother \do\$\do\&\do\#\do\^\do\_\do\%\do\~}
\def\mn@doi{\begingroup\mn@urlcharsother \@ifnextchar [ {\mn@doi@}
  {\mn@doi@[]}}
\def\mn@doi@[#1]#2{\def\@tempa{#1}\ifx\@tempa\@empty \href
  {http://dx.doi.org/#2} {doi:#2}\else \href {http://dx.doi.org/#2} {#1}\fi
  \endgroup}
\def\mn@eprint#1#2{\mn@eprint@#1:#2::\@nil}
\def\mn@eprint@arXiv#1{\href {http://arxiv.org/abs/#1} {{\tt arXiv:#1}}}
\def\mn@eprint@dblp#1{\href {http://dblp.uni-trier.de/rec/bibtex/#1.xml}
  {dblp:#1}}
\def\mn@eprint@#1:#2:#3:#4\@nil{\def\@tempa {#1}\def\@tempb {#2}\def\@tempc
  {#3}\ifx \@tempc \@empty \let \@tempc \@tempb \let \@tempb \@tempa \fi \ifx
  \@tempb \@empty \def\@tempb {arXiv}\fi \@ifundefined
  {mn@eprint@\@tempb}{\@tempb:\@tempc}{\expandafter \expandafter \csname
  mn@eprint@\@tempb\endcsname \expandafter{\@tempc}}}

\bibitem[\protect\citeauthoryear{{Allain}}{{Allain}}{1998}]{1998A&A...333..629A}
{Allain} S.,  1998, \aap, \href
  {https://ui.adsabs.harvard.edu/abs/1998A&A...333..629A} {333, 629}

\bibitem[\protect\citeauthoryear{{Amard}, {Palacios}, {Charbonnel}, {Gallet},
  {Georgy}, {Lagarde}  \& {Siess}}{{Amard} et~al.}{2019}]{2019A&A...631A..77A}
{Amard} L.,  {Palacios} A.,  {Charbonnel} C.,  {Gallet} F.,  {Georgy} C.,
  {Lagarde} N.,   {Siess} L.,  2019, \mn@doi [\aap]
  {10.1051/0004-6361/201935160}, \href
  {https://ui.adsabs.harvard.edu/abs/2019A&A...631A..77A} {631, A77}

\bibitem[\protect\citeauthoryear{{Baraffe} \& {Chabrier}}{{Baraffe} \&
  {Chabrier}}{2010}]{2010A&A...521A..44B}
{Baraffe} I.,  {Chabrier} G.,  2010, \mn@doi [\aap]
  {10.1051/0004-6361/201014979}, \href
  {https://ui.adsabs.harvard.edu/abs/2010A&A...521A..44B} {521, A44}

\bibitem[\protect\citeauthoryear{{Baraffe}, {Chabrier}, {Allard}  \&
  {Hauschildt}}{{Baraffe} et~al.}{1998}]{1998A&A...337..403B}
{Baraffe} I.,  {Chabrier} G.,  {Allard} F.,   {Hauschildt} P.~H.,  1998, \aap,
  \href {http://adsabs.harvard.edu/abs/1998A%26A...337..403B} {337, 403}

\bibitem[\protect\citeauthoryear{{Barnes}}{{Barnes}}{2010}]{2010ApJ...722..222B}
{Barnes} S.~A.,  2010, \mn@doi [\apj] {10.1088/0004-637X/722/1/222}, \href
  {https://ui.adsabs.harvard.edu/abs/2010ApJ...722..222B} {722, 222}

\bibitem[\protect\citeauthoryear{{Barnes} \& {Kim}}{{Barnes} \&
  {Kim}}{2010}]{2010ApJ...721..675B}
{Barnes} S.~A.,  {Kim} Y.-C.,  2010, \mn@doi [\apj]
  {10.1088/0004-637X/721/1/675}, \href
  {https://ui.adsabs.harvard.edu/abs/2010ApJ...721..675B} {721, 675}

\bibitem[\protect\citeauthoryear{{Bastian}, {Covey}  \& {Meyer}}{{Bastian}
  et~al.}{2010}]{2010ARA&A..48..339B}
{Bastian} N.,  {Covey} K.~R.,   {Meyer} M.~R.,  2010, \mn@doi [\araa]
  {10.1146/annurev-astro-082708-101642}, \href
  {https://ui.adsabs.harvard.edu/abs/2010ARA&A..48..339B} {48, 339}

\bibitem[\protect\citeauthoryear{{Berlanas}, {Herrero}, {Comer{\'o}n},
  {Sim{\'o}n-D{\'\i}az}, {Cervi{\~n}o}  \& {Pasquali}}{{Berlanas}
  et~al.}{2018}]{2018A&A...620A..56B}
{Berlanas} S.~R.,  {Herrero} A.,  {Comer{\'o}n} F.,  {Sim{\'o}n-D{\'\i}az} S.,
  {Cervi{\~n}o} M.,   {Pasquali} A.,  2018, \mn@doi [\aap]
  {10.1051/0004-6361/201833989}, \href
  {https://ui.adsabs.harvard.edu/abs/2018A&A...620A..56B} {620, A56}

\bibitem[\protect\citeauthoryear{{Biazzo}, {Randich}  \& {Palla}}{{Biazzo}
  et~al.}{2011}]{2011A&A...525A..35B}
{Biazzo} K.,  {Randich} S.,   {Palla} F.,  2011, \mn@doi [\aap]
  {10.1051/0004-6361/201015489}, \href
  {https://ui.adsabs.harvard.edu/abs/2011A&A...525A..35B} {525, A35}

\bibitem[\protect\citeauthoryear{{Bouvier}, {Matt}, {Mohanty}, {Scholz},
  {Stassun}  \& {Zanni}}{{Bouvier} et~al.}{2014}]{2014prpl.conf..433B}
{Bouvier} J.,  {Matt} S.~P.,  {Mohanty} S.,  {Scholz} A.,  {Stassun} K.~G.,
  {Zanni} C.,  2014, \mn@doi [Protostars and Planets VI]
  {10.2458/azu_uapress_9780816531240-ch019}, \href
  {http://adsabs.harvard.edu/abs/2014prpl.conf..433B} {pp 433--450}

\bibitem[\protect\citeauthoryear{{Casamiquela} et~al.,}{{Casamiquela}
  et~al.}{2017}]{2017MNRAS.470.4363C}
{Casamiquela} L.,  et~al., 2017, \mn@doi [\mnras] {10.1093/mnras/stx1481},
  \href {https://ui.adsabs.harvard.edu/abs/2017MNRAS.470.4363C} {470, 4363}

\bibitem[\protect\citeauthoryear{{Chaboyer}, {Demarque}  \&
  {Pinsonneault}}{{Chaboyer} et~al.}{1995}]{1995ApJ...441..876C}
{Chaboyer} B.,  {Demarque} P.,   {Pinsonneault} M.~H.,  1995, \mn@doi [\apj]
  {10.1086/175409}, \href
  {https://ui.adsabs.harvard.edu/abs/1995ApJ...441..876C} {441, 876}

\bibitem[\protect\citeauthoryear{{Cieza} \& {Baliber}}{{Cieza} \&
  {Baliber}}{2007}]{2007ApJ...671..605C}
{Cieza} L.,  {Baliber} N.,  2007, \mn@doi [\apj] {10.1086/522080}, \href
  {http://adsabs.harvard.edu/abs/2007ApJ...671..605C} {671, 605}

\bibitem[\protect\citeauthoryear{{Coker}, {Pinsonneault}  \&
  {Terndrup}}{{Coker} et~al.}{2016}]{2016ApJ...833..122C}
{Coker} C.~T.,  {Pinsonneault} M.,   {Terndrup} D.~M.,  2016, \mn@doi [\apj]
  {10.3847/1538-4357/833/1/122}, \href
  {https://ui.adsabs.harvard.edu/abs/2016ApJ...833..122C} {833, 122}

\bibitem[\protect\citeauthoryear{{Concha-Ram{\'\i}rez}, {Wilhelm}, {Portegies
  Zwart}, {van Terwisga}  \& {Hacar}}{{Concha-Ram{\'\i}rez}
  et~al.}{2021}]{2021MNRAS.501.1782C}
{Concha-Ram{\'\i}rez} F.,  {Wilhelm} M. J.~C.,  {Portegies Zwart} S.,  {van
  Terwisga} S.~E.,   {Hacar} A.,  2021, \mn@doi [\mnras]
  {10.1093/mnras/staa3669}, \href
  {https://ui.adsabs.harvard.edu/abs/2021MNRAS.501.1782C} {501, 1782}

\bibitem[\protect\citeauthoryear{{Covey} et~al.,}{{Covey}
  et~al.}{2016}]{2016ApJ...822...81C}
{Covey} K.~R.,  et~al., 2016, \mn@doi [\apj] {10.3847/0004-637X/822/2/81},
  \href {https://ui.adsabs.harvard.edu/abs/2016ApJ...822...81C} {822, 81}

\bibitem[\protect\citeauthoryear{{Cranmer} \& {Saar}}{{Cranmer} \&
  {Saar}}{2011}]{2011ApJ...741...54C}
{Cranmer} S.~R.,  {Saar} S.~H.,  2011, \mn@doi [\apj]
  {10.1088/0004-637X/741/1/54}, \href
  {https://ui.adsabs.harvard.edu/abs/2011ApJ...741...54C} {741, 54}

\bibitem[\protect\citeauthoryear{{D'Orazi}, {Randich}, {Flaccomio}, {Palla},
  {Sacco}  \& {Pallavicini}}{{D'Orazi} et~al.}{2009}]{2009A&A...501..973D}
{D'Orazi} V.,  {Randich} S.,  {Flaccomio} E.,  {Palla} F.,  {Sacco} G.~G.,
  {Pallavicini} R.,  2009, \mn@doi [\aap] {10.1051/0004-6361/200811241}, \href
  {https://ui.adsabs.harvard.edu/abs/2009A&A...501..973D} {501, 973}

\bibitem[\protect\citeauthoryear{{Damian}, {Jose}, {Samal}, {Moraux}, {Das}  \&
  {Patra}}{{Damian} et~al.}{2021}]{2021arXiv210108804D}
{Damian} B.,  {Jose} J.,  {Samal} M.~R.,  {Moraux} E.,  {Das} S.~R.,   {Patra}
  S.,  2021, arXiv e-prints, \href
  {https://ui.adsabs.harvard.edu/abs/2021arXiv210108804D} {p. arXiv:2101.08804}

\bibitem[\protect\citeauthoryear{{Davies}, {Gregory}  \& {Greaves}}{{Davies}
  et~al.}{2014}]{2014MNRAS.444.1157D}
{Davies} C.~L.,  {Gregory} S.~G.,   {Greaves} J.~S.,  2014, \mn@doi [\mnras]
  {10.1093/mnras/stu1488}, \href
  {http://adsabs.harvard.edu/abs/2014MNRAS.444.1157D} {444, 1157}

\bibitem[\protect\citeauthoryear{{Demarque}, {Guenther}, {Li}, {Mazumdar}  \&
  {Straka}}{{Demarque} et~al.}{2008}]{2008Ap&SS.316...31D}
{Demarque} P.,  {Guenther} D.~B.,  {Li} L.~H.,  {Mazumdar} A.,   {Straka}
  C.~W.,  2008, \mn@doi [\apss] {10.1007/s10509-007-9698-y}, \href
  {https://ui.adsabs.harvard.edu/abs/2008Ap&SS.316...31D} {316, 31}

\bibitem[\protect\citeauthoryear{{Dufton}, {Brown}, {Fitzsimmons}  \&
  {Lennon}}{{Dufton} et~al.}{1990}]{1990A&A...232..431D}
{Dufton} P.~L.,  {Brown} P.~J.~F.,  {Fitzsimmons} A.,   {Lennon} D.~J.,  1990,
  \aap, \href {https://ui.adsabs.harvard.edu/abs/1990A&A...232..431D} {232,
  431}

\bibitem[\protect\citeauthoryear{{Edwards} et~al.,}{{Edwards}
  et~al.}{1993}]{1993AJ....106..372E}
{Edwards} S.,  et~al., 1993, \mn@doi [\aj] {10.1086/116646}, \href
  {https://ui.adsabs.harvard.edu/abs/1993AJ....106..372E} {106, 372}

\bibitem[\protect\citeauthoryear{{Ferreira}, {Pelletier}  \& {Appl}}{{Ferreira}
  et~al.}{2000}]{2000MNRAS.312..387F}
{Ferreira} J.,  {Pelletier} G.,   {Appl} S.,  2000, \mn@doi [\mnras]
  {10.1046/j.1365-8711.2000.03215.x}, \href
  {https://ui.adsabs.harvard.edu/abs/2000MNRAS.312..387F} {312, 387}

\bibitem[\protect\citeauthoryear{{Gallet} \& {Bouvier}}{{Gallet} \&
  {Bouvier}}{2013}]{2013A&A...556A..36G}
{Gallet} F.,  {Bouvier} J.,  2013, \mn@doi [\aap]
  {10.1051/0004-6361/201321302}, \href
  {http://adsabs.harvard.edu/abs/2013A%26A...556A..36G} {556, A36}

\bibitem[\protect\citeauthoryear{{Gallet} \& {Bouvier}}{{Gallet} \&
  {Bouvier}}{2015}]{2015A&A...577A..98G}
{Gallet} F.,  {Bouvier} J.,  2015, \mn@doi [\aap]
  {10.1051/0004-6361/201525660}, \href
  {http://adsabs.harvard.edu/abs/2015A%26A...577A..98G} {577, A98}

\bibitem[\protect\citeauthoryear{{Gallet}, {Zanni}  \& {Amard}}{{Gallet}
  et~al.}{2019}]{2019A&A...632A...6G}
{Gallet} F.,  {Zanni} C.,   {Amard} L.,  2019, \mn@doi [\aap]
  {10.1051/0004-6361/201935432}, \href
  {https://ui.adsabs.harvard.edu/abs/2019A&A...632A...6G} {632, A6}

\bibitem[\protect\citeauthoryear{{Ghosh} \& {Lamb}}{{Ghosh} \&
  {Lamb}}{1979}]{1979ApJ...234..296G}
{Ghosh} P.,  {Lamb} F.~K.,  1979, \mn@doi [\apj] {10.1086/157498}, \href
  {https://ui.adsabs.harvard.edu/abs/1979ApJ...234..296G} {234, 296}

\bibitem[\protect\citeauthoryear{{Gondoin}}{{Gondoin}}{2017}]{2017A&A...599A.122G}
{Gondoin} P.,  2017, \mn@doi [\aap] {10.1051/0004-6361/201629760}, \href
  {https://ui.adsabs.harvard.edu/abs/2017A&A...599A.122G} {599, A122}

\bibitem[\protect\citeauthoryear{{Guarcello} et~al.,}{{Guarcello}
  et~al.}{2013}]{2013ApJ...773..135G}
{Guarcello} M.~G.,  et~al., 2013, \mn@doi [\apj] {10.1088/0004-637X/773/2/135},
  \href {https://ui.adsabs.harvard.edu/abs/2013ApJ...773..135G} {773, 135}

\bibitem[\protect\citeauthoryear{{Guarcello} et~al.,}{{Guarcello}
  et~al.}{2021}]{2021A&A...650A.157G}
{Guarcello} M.~G.,  et~al., 2021, \mn@doi [\aap] {10.1051/0004-6361/202140361},
  \href {https://ui.adsabs.harvard.edu/abs/2021A&A...650A.157G} {650, A157}

\bibitem[\protect\citeauthoryear{{Hartman} et~al.,}{{Hartman}
  et~al.}{2008}]{2008ApJ...675.1233H}
{Hartman} J.~D.,  et~al., 2008, \mn@doi [\apj] {10.1086/527465}, \href
  {https://ui.adsabs.harvard.edu/abs/2008ApJ...675.1233H} {675, 1233}

\bibitem[\protect\citeauthoryear{{Hartman} et~al.,}{{Hartman}
  et~al.}{2009}]{2009ApJ...691..342H}
{Hartman} J.~D.,  et~al., 2009, \mn@doi [\apj] {10.1088/0004-637X/691/1/342},
  \href {https://ui.adsabs.harvard.edu/abs/2009ApJ...691..342H} {691, 342}

\bibitem[\protect\citeauthoryear{{Hartman}, {Bakos}, {Kov{\'a}cs}  \&
  {Noyes}}{{Hartman} et~al.}{2010}]{2010MNRAS.408..475H}
{Hartman} J.~D.,  {Bakos} G.~{\'A}.,  {Kov{\'a}cs} G.,   {Noyes} R.~W.,  2010,
  \mn@doi [\mnras] {10.1111/j.1365-2966.2010.17147.x}, \href
  {https://ui.adsabs.harvard.edu/abs/2010MNRAS.408..475H} {408, 475}

\bibitem[\protect\citeauthoryear{{Hartmann}, {Calvet}, {Gullbring}  \&
  {D'Alessio}}{{Hartmann} et~al.}{1998}]{1998ApJ...495..385H}
{Hartmann} L.,  {Calvet} N.,  {Gullbring} E.,   {D'Alessio} P.,  1998, \mn@doi
  [\apj] {10.1086/305277}, \href
  {https://ui.adsabs.harvard.edu/abs/1998ApJ...495..385H} {495, 385}

\bibitem[\protect\citeauthoryear{{Herczeg} \& {Hillenbrand}}{{Herczeg} \&
  {Hillenbrand}}{2015}]{2015ApJ...808...23H}
{Herczeg} G.~J.,  {Hillenbrand} L.~A.,  2015, \mn@doi [\apj]
  {10.1088/0004-637X/808/1/23}, \href
  {https://ui.adsabs.harvard.edu/abs/2015ApJ...808...23H} {808, 23}

\bibitem[\protect\citeauthoryear{{Hern{\'a}ndez} et~al.,}{{Hern{\'a}ndez}
  et~al.}{2007}]{2007ApJ...662.1067H}
{Hern{\'a}ndez} J.,  et~al., 2007, \mn@doi [\apj] {10.1086/513735}, \href
  {http://adsabs.harvard.edu/abs/2007ApJ...662.1067H} {662, 1067}

\bibitem[\protect\citeauthoryear{{Hern{\'a}ndez}, {Hartmann}, {Calvet},
  {Jeffries}, {Gutermuth}, {Muzerolle}  \& {Stauffer}}{{Hern{\'a}ndez}
  et~al.}{2008}]{2008ApJ...686.1195H}
{Hern{\'a}ndez} J.,  {Hartmann} L.,  {Calvet} N.,  {Jeffries} R.~D.,
  {Gutermuth} R.,  {Muzerolle} J.,   {Stauffer} J.,  2008, \mn@doi [\apj]
  {10.1086/591224}, \href {http://adsabs.harvard.edu/abs/2008ApJ...686.1195H}
  {686, 1195}

\bibitem[\protect\citeauthoryear{{Ireland}, {Zanni}, {Matt}  \&
  {Pantolmos}}{{Ireland} et~al.}{2021}]{2021ApJ...906....4I}
{Ireland} L.~G.,  {Zanni} C.,  {Matt} S.~P.,   {Pantolmos} G.,  2021, \mn@doi
  [\apj] {10.3847/1538-4357/abc828}, \href
  {https://ui.adsabs.harvard.edu/abs/2021ApJ...906....4I} {906, 4}

\bibitem[\protect\citeauthoryear{{Irwin}, {Hodgkin}, {Aigrain}, {Hebb},
  {Bouvier}, {Clarke}, {Moraux}  \& {Bramich}}{{Irwin}
  et~al.}{2007}]{2007MNRAS.377..741I}
{Irwin} J.,  {Hodgkin} S.,  {Aigrain} S.,  {Hebb} L.,  {Bouvier} J.,  {Clarke}
  C.,  {Moraux} E.,   {Bramich} D.~M.,  2007, \mn@doi [\mnras]
  {10.1111/j.1365-2966.2007.11640.x}, \href
  {https://ui.adsabs.harvard.edu/abs/2007MNRAS.377..741I} {377, 741}

\bibitem[\protect\citeauthoryear{{Irwin}, {Hodgkin}, {Aigrain}, {Bouvier},
  {Hebb}, {Irwin}  \& {Moraux}}{{Irwin} et~al.}{2008}]{2008MNRAS.384..675I}
{Irwin} J.,  {Hodgkin} S.,  {Aigrain} S.,  {Bouvier} J.,  {Hebb} L.,  {Irwin}
  M.,   {Moraux} E.,  2008, \mn@doi [\mnras]
  {10.1111/j.1365-2966.2007.12725.x}, \href
  {https://ui.adsabs.harvard.edu/abs/2008MNRAS.384..675I} {384, 675}

\bibitem[\protect\citeauthoryear{{Kawaler}}{{Kawaler}}{1988}]{1988ApJ...333..236K}
{Kawaler} S.~D.,  1988, \mn@doi [\apj] {10.1086/166740}, \href
  {https://ui.adsabs.harvard.edu/abs/1988ApJ...333..236K} {333, 236}

\bibitem[\protect\citeauthoryear{{Koenigl}}{{Koenigl}}{1991}]{1991ApJ...370L..39K}
{Koenigl} A.,  1991, \mn@doi [\apjl] {10.1086/185972}, \href
  {http://adsabs.harvard.edu/abs/1991ApJ...370L..39K} {370, L39}

\bibitem[\protect\citeauthoryear{{Kroupa}, {Weidner}, {Pflamm-Altenburg},
  {Thies}, {Dabringhausen}, {Marks}  \& {Maschberger}}{{Kroupa}
  et~al.}{2013}]{2013pss5.book..115K}
{Kroupa} P.,  {Weidner} C.,  {Pflamm-Altenburg} J.,  {Thies} I.,
  {Dabringhausen} J.,  {Marks} M.,   {Maschberger} T.,  2013, {The Stellar and
  Sub-Stellar Initial Mass Function of Simple and Composite Populations}.
p.~115, \mn@doi{10.1007/978-94-007-5612-0_4}

\bibitem[\protect\citeauthoryear{{Landin}, {Mendes}, {Vaz}  \&
  {Alencar}}{{Landin} et~al.}{2016}]{2016A&A...586A..96L}
{Landin} N.~R.,  {Mendes} L.~T.~S.,  {Vaz} L.~P.~R.,   {Alencar} S.~H.~P.,
  2016, \mn@doi [\aap] {10.1051/0004-6361/201525851}, \href
  {https://ui.adsabs.harvard.edu/abs/2016A&A...586A..96L} {586, A96}

\bibitem[\protect\citeauthoryear{{Lanzafame} \& {Spada}}{{Lanzafame} \&
  {Spada}}{2015}]{2015A&A...584A..30L}
{Lanzafame} A.~C.,  {Spada} F.,  2015, \mn@doi [\aap]
  {10.1051/0004-6361/201526770}, \href
  {https://ui.adsabs.harvard.edu/abs/2015A&A...584A..30L} {584, A30}

\bibitem[\protect\citeauthoryear{{Littlefair}, {Naylor}, {Mayne}, {Saunders}
  \& {Jeffries}}{{Littlefair} et~al.}{2010}]{2010MNRAS.403..545L}
{Littlefair} S.~P.,  {Naylor} T.,  {Mayne} N.~J.,  {Saunders} E.~S.,
  {Jeffries} R.~D.,  2010, \mn@doi [\mnras] {10.1111/j.1365-2966.2010.16066.x},
  \href {https://ui.adsabs.harvard.edu/abs/2010MNRAS.403..545L} {403, 545}

\bibitem[\protect\citeauthoryear{{Lodieu}, {P{\'e}rez-Garrido}, {Smart}  \&
  {Silvotti}}{{Lodieu} et~al.}{2019}]{2019A&A...628A..66L}
{Lodieu} N.,  {P{\'e}rez-Garrido} A.,  {Smart} R.~L.,   {Silvotti} R.,  2019,
  \mn@doi [\aap] {10.1051/0004-6361/201935533}, \href
  {https://ui.adsabs.harvard.edu/abs/2019A&A...628A..66L} {628, A66}

\bibitem[\protect\citeauthoryear{{MacGregor} \& {Brenner}}{{MacGregor} \&
  {Brenner}}{1991}]{1991ApJ...376..204M}
{MacGregor} K.~B.,  {Brenner} M.,  1991, \mn@doi [\apj] {10.1086/170269}, \href
  {https://ui.adsabs.harvard.edu/abs/1991ApJ...376..204M} {376, 204}

\bibitem[\protect\citeauthoryear{{Matt} \& {Pudritz}}{{Matt} \&
  {Pudritz}}{2005}]{2005ApJ...632L.135M}
{Matt} S.,  {Pudritz} R.~E.,  2005, \mn@doi [\apjl] {10.1086/498066}, \href
  {https://ui.adsabs.harvard.edu/abs/2005ApJ...632L.135M} {632, L135}

\bibitem[\protect\citeauthoryear{{Matt}, {MacGregor}, {Pinsonneault}  \&
  {Greene}}{{Matt} et~al.}{2012}]{2012ApJ...754L..26M}
{Matt} S.~P.,  {MacGregor} K.~B.,  {Pinsonneault} M.~H.,   {Greene} T.~P.,
  2012, \mn@doi [\apjl] {10.1088/2041-8205/754/2/L26}, \href
  {https://ui.adsabs.harvard.edu/abs/2012ApJ...754L..26M} {754, L26}

\bibitem[\protect\citeauthoryear{{Matt}, {Brun}, {Baraffe}, {Bouvier}  \&
  {Chabrier}}{{Matt} et~al.}{2015}]{2015ApJ...799L..23M}
{Matt} S.~P.,  {Brun} A.~S.,  {Baraffe} I.,  {Bouvier} J.,   {Chabrier} G.,
  2015, \mn@doi [\apjl] {10.1088/2041-8205/799/2/L23}, \href
  {https://ui.adsabs.harvard.edu/abs/2015ApJ...799L..23M} {799, L23}

\bibitem[\protect\citeauthoryear{{Mayne} \& {Naylor}}{{Mayne} \&
  {Naylor}}{2008}]{2008MNRAS.386..261M}
{Mayne} N.~J.,  {Naylor} T.,  2008, \mn@doi [\mnras]
  {10.1111/j.1365-2966.2008.13025.x}, \href
  {https://ui.adsabs.harvard.edu/abs/2008MNRAS.386..261M} {386, 261}

\bibitem[\protect\citeauthoryear{{Moraux} et~al.,}{{Moraux}
  et~al.}{2013}]{2013A&A...560A..13M}
{Moraux} E.,  et~al., 2013, \mn@doi [\aap] {10.1051/0004-6361/201321508}, \href
  {http://adsabs.harvard.edu/abs/2013A%26A...560A..13M} {560, A13}

\bibitem[\protect\citeauthoryear{{Netopil}, {Paunzen}, {Heiter}  \&
  {Soubiran}}{{Netopil} et~al.}{2016}]{2016A&A...585A.150N}
{Netopil} M.,  {Paunzen} E.,  {Heiter} U.,   {Soubiran} C.,  2016, \mn@doi
  [\aap] {10.1051/0004-6361/201526370}, \href
  {https://ui.adsabs.harvard.edu/abs/2016A&A...585A.150N} {585, A150}

\bibitem[\protect\citeauthoryear{{Padgett}}{{Padgett}}{1996}]{1996ApJ...471..847P}
{Padgett} D.~L.,  1996, \mn@doi [\apj] {10.1086/178012}, \href
  {https://ui.adsabs.harvard.edu/abs/1996ApJ...471..847P} {471, 847}

\bibitem[\protect\citeauthoryear{{Pecaut}, {Mamajek}  \& {Bubar}}{{Pecaut}
  et~al.}{2012}]{2012ApJ...746..154P}
{Pecaut} M.~J.,  {Mamajek} E.~E.,   {Bubar} E.~J.,  2012, \mn@doi [\apj]
  {10.1088/0004-637X/746/2/154}, \href
  {https://ui.adsabs.harvard.edu/abs/2012ApJ...746..154P} {746, 154}

\bibitem[\protect\citeauthoryear{{Rebull}, {Wolff}  \& {Strom}}{{Rebull}
  et~al.}{2004}]{2004AJ....127.1029R}
{Rebull} L.~M.,  {Wolff} S.~C.,   {Strom} S.~E.,  2004, \mn@doi [\aj]
  {10.1086/380931}, \href
  {https://ui.adsabs.harvard.edu/abs/2004AJ....127.1029R} {127, 1029}

\bibitem[\protect\citeauthoryear{{Rebull}, {Stauffer}, {Megeath}, {Hora}  \&
  {Hartmann}}{{Rebull} et~al.}{2006}]{2006ApJ...646..297R}
{Rebull} L.~M.,  {Stauffer} J.~R.,  {Megeath} S.~T.,  {Hora} J.~L.,
  {Hartmann} L.,  2006, \mn@doi [\apj] {10.1086/504865}, \href
  {https://ui.adsabs.harvard.edu/abs/2006ApJ...646..297R} {646, 297}

\bibitem[\protect\citeauthoryear{{Rebull} et~al.,}{{Rebull}
  et~al.}{2016}]{2016AJ....152..113R}
{Rebull} L.~M.,  et~al., 2016, \mn@doi [\aj] {10.3847/0004-6256/152/5/113},
  \href {https://ui.adsabs.harvard.edu/abs/2016AJ....152..113R} {152, 113}

\bibitem[\protect\citeauthoryear{{Rebull}, {Stauffer}, {Hillenbrand}, {Cody},
  {Bouvier}, {Soderblom}, {Pinsonneault}  \& {Hebb}}{{Rebull}
  et~al.}{2017}]{2017ApJ...839...92R}
{Rebull} L.~M.,  {Stauffer} J.~R.,  {Hillenbrand} L.~A.,  {Cody} A.~M.,
  {Bouvier} J.,  {Soderblom} D.~R.,  {Pinsonneault} M.,   {Hebb} L.,  2017,
  \mn@doi [\apj] {10.3847/1538-4357/aa6aa4}, \href
  {https://ui.adsabs.harvard.edu/abs/2017ApJ...839...92R} {839, 92}

\bibitem[\protect\citeauthoryear{{Rebull}, {Stauffer}, {Cody}, {Hillenbrand },
  {David}  \& {Pinsonneault}}{{Rebull} et~al.}{2018}]{2018AJ....155..196R}
{Rebull} L.~M.,  {Stauffer} J.~R.,  {Cody} A.~M.,  {Hillenbrand } L.~A.,
  {David} T.~J.,   {Pinsonneault} M.,  2018, \mn@doi [\aj]
  {10.3847/1538-3881/aab605}, \href
  {https://ui.adsabs.harvard.edu/abs/2018AJ....155..196R} {155, 196}

\bibitem[\protect\citeauthoryear{{Rebull}, {Stauffer}, {Cody}, {Hillenbrand },
  {Bouvier}, {Roggero}  \& {David}}{{Rebull}
  et~al.}{2020}]{2020AJ....159..273R}
{Rebull} L.~M.,  {Stauffer} J.~R.,  {Cody} A.~M.,  {Hillenbrand } L.~A.,
  {Bouvier} J.,  {Roggero} N.,   {David} T.~J.,  2020, \mn@doi [\aj]
  {10.3847/1538-3881/ab893c}, \href
  {https://ui.adsabs.harvard.edu/abs/2020AJ....159..273R} {159, 273}

\bibitem[\protect\citeauthoryear{{Ribas}, {Mer{\'{\i}}n}, {Bouy}  \&
  {Maud}}{{Ribas} et~al.}{2014}]{2014A&A...561A..54R}
{Ribas} {\'A}.,  {Mer{\'{\i}}n} B.,  {Bouy} H.,   {Maud} L.~T.,  2014, \mn@doi
  [\aap] {10.1051/0004-6361/201322597}, \href
  {http://adsabs.harvard.edu/abs/2014A%26A...561A..54R} {561, A54}

\bibitem[\protect\citeauthoryear{{Romanova}, {Ustyugova}, {Koldoba}  \&
  {Lovelace}}{{Romanova} et~al.}{2009}]{2009MNRAS.399.1802R}
{Romanova} M.~M.,  {Ustyugova} G.~V.,  {Koldoba} A.~V.,   {Lovelace} R.~V.~E.,
  2009, \mn@doi [\mnras] {10.1111/j.1365-2966.2009.15413.x}, \href
  {https://ui.adsabs.harvard.edu/abs/2009MNRAS.399.1802R} {399, 1802}

\bibitem[\protect\citeauthoryear{{Roquette}, {Bouvier}, {Alencar}, {Vaz}  \&
  {Guarcello}}{{Roquette} et~al.}{2017}]{2017A&A...603A.106R}
{Roquette} J.,  {Bouvier} J.,  {Alencar} S.~H.~P.,  {Vaz} L.~P.~R.,
  {Guarcello} M.~G.,  2017, \mn@doi [\aap] {10.1051/0004-6361/201630337}, \href
  {https://ui.adsabs.harvard.edu/abs/2017A&A...603A.106R} {603, A106}

\bibitem[\protect\citeauthoryear{{Roquette}, {Matt}, {Winter}, {Amard}  \&
  {Stasevic}}{{Roquette} et~al.}{2021}]{2021MNRAS.508.3710R}
{Roquette} J.,  {Matt} S.~P.,  {Winter} A.~J.,  {Amard} L.,   {Stasevic} S.,
  2021, \mn@doi [\mnras] {10.1093/mnras/stab2772}, \href
  {https://ui.adsabs.harvard.edu/abs/2021MNRAS.508.3710R} {508, 3710}

\bibitem[\protect\citeauthoryear{{Shu}, {Najita}, {Ostriker}, {Wilkin}, {Ruden}
   \& {Lizano}}{{Shu} et~al.}{1994}]{1994ApJ...429..781S}
{Shu} F.,  {Najita} J.,  {Ostriker} E.,  {Wilkin} F.,  {Ruden} S.,   {Lizano}
  S.,  1994, \mn@doi [\apj] {10.1086/174363}, \href
  {https://ui.adsabs.harvard.edu/abs/1994ApJ...429..781S} {429, 781}

\bibitem[\protect\citeauthoryear{{Slesnick}, {Hillenbrand}  \&
  {Carpenter}}{{Slesnick} et~al.}{2008}]{2008ApJ...688..377S}
{Slesnick} C.~L.,  {Hillenbrand} L.~A.,   {Carpenter} J.~M.,  2008, \mn@doi
  [\apj] {10.1086/592265}, \href
  {https://ui.adsabs.harvard.edu/abs/2008ApJ...688..377S} {688, 377}

\bibitem[\protect\citeauthoryear{{Smartt} \& {Rolleston}}{{Smartt} \&
  {Rolleston}}{1997}]{1997ApJ...481L..47S}
{Smartt} S.~J.,  {Rolleston} W. R.~J.,  1997, \mn@doi [\apjl] {10.1086/310640},
  \href {https://ui.adsabs.harvard.edu/abs/1997ApJ...481L..47S} {481, L47}

\bibitem[\protect\citeauthoryear{{Soderblom}, {Hillenbrand}, {Jeffries},
  {Mamajek}  \& {Naylor}}{{Soderblom} et~al.}{2014}]{2014prpl.conf..219S}
{Soderblom} D.~R.,  {Hillenbrand} L.~A.,  {Jeffries} R.~D.,  {Mamajek} E.~E.,
  {Naylor} T.,  2014, in {Beuther} H.,  {Klessen} R.~S.,  {Dullemond} C.~P.,
  {Henning} T.,  eds, Protostars and Planets VI. p.~219 (\mn@eprint {arXiv}
  {1311.7024}), \mn@doi{10.2458/azu\_uapress\_9780816531240-ch010}

\bibitem[\protect\citeauthoryear{{Somers} \& {Pinsonneault}}{{Somers} \&
  {Pinsonneault}}{2015}]{2015MNRAS.449.4131S}
{Somers} G.,  {Pinsonneault} M.~H.,  2015, \mn@doi [\mnras]
  {10.1093/mnras/stv630}, \href
  {https://ui.adsabs.harvard.edu/abs/2015MNRAS.449.4131S} {449, 4131}

\bibitem[\protect\citeauthoryear{{Southworth}, {Maxted}  \&
  {Smalley}}{{Southworth} et~al.}{2004}]{2004MNRAS.349..547S}
{Southworth} J.,  {Maxted} P.~F.~L.,   {Smalley} B.,  2004, \mn@doi [\mnras]
  {10.1111/j.1365-2966.2004.07520.x}, \href
  {https://ui.adsabs.harvard.edu/abs/2004MNRAS.349..547S} {349, 547}

\bibitem[\protect\citeauthoryear{{Spada} \& {Lanzafame}}{{Spada} \&
  {Lanzafame}}{2020}]{2020A&A...636A..76S}
{Spada} F.,  {Lanzafame} A.~C.,  2020, \mn@doi [\aap]
  {10.1051/0004-6361/201936384}, \href
  {https://ui.adsabs.harvard.edu/abs/2020A&A...636A..76S} {636, A76}

\bibitem[\protect\citeauthoryear{{Spada}, {Lanzafame}, {Lanza}, {Messina}  \&
  {Collier Cameron}}{{Spada} et~al.}{2011}]{2011MNRAS.416..447S}
{Spada} F.,  {Lanzafame} A.~C.,  {Lanza} A.~F.,  {Messina} S.,   {Collier
  Cameron} A.,  2011, \mn@doi [\mnras] {10.1111/j.1365-2966.2011.19052.x},
  \href {https://ui.adsabs.harvard.edu/abs/2011MNRAS.416..447S} {416, 447}

\bibitem[\protect\citeauthoryear{{Stauffer}, {Schultz}  \&
  {Kirkpatrick}}{{Stauffer} et~al.}{1998}]{1998ApJ...499L.199S}
{Stauffer} J.~R.,  {Schultz} G.,   {Kirkpatrick} J.~D.,  1998, \mn@doi [\apjl]
  {10.1086/311379}, \href
  {https://ui.adsabs.harvard.edu/abs/1998ApJ...499L.199S} {499, L199}

\bibitem[\protect\citeauthoryear{{Takeda}, {Hashimoto}  \& {Honda}}{{Takeda}
  et~al.}{2017}]{2017PASJ...69....1T}
{Takeda} Y.,  {Hashimoto} O.,   {Honda} S.,  2017, \mn@doi [\pasj]
  {10.1093/pasj/psw105}, \href
  {https://ui.adsabs.harvard.edu/abs/2017PASJ...69....1T} {69, 1}

\bibitem[\protect\citeauthoryear{{Tamajo}, {Pavlovski}  \&
  {Southworth}}{{Tamajo} et~al.}{2011}]{2011A&A...526A..76T}
{Tamajo} E.,  {Pavlovski} K.,   {Southworth} J.,  2011, \mn@doi [\aap]
  {10.1051/0004-6361/201015913}, \href
  {https://ui.adsabs.harvard.edu/abs/2011A&A...526A..76T} {526, A76}

\bibitem[\protect\citeauthoryear{{Vasconcelos} \& {Bouvier}}{{Vasconcelos} \&
  {Bouvier}}{2015}]{2015A&A...578A..89V}
{Vasconcelos} M.~J.,  {Bouvier} J.,  2015, \mn@doi [\aap]
  {10.1051/0004-6361/201525765}, \href
  {http://adsabs.harvard.edu/abs/2015A%26A...578A..89V} {578, A89}

\bibitem[\protect\citeauthoryear{{Vasconcelos} \& {Bouvier}}{{Vasconcelos} \&
  {Bouvier}}{2017}]{2017A&A...600A.116V}
{Vasconcelos} M.~J.,  {Bouvier} J.,  2017, \mn@doi [\aap]
  {10.1051/0004-6361/201628724}, \href
  {https://ui.adsabs.harvard.edu/abs/2017A&A...600A.116V} {600, A116}

\bibitem[\protect\citeauthoryear{{Viana Almeida}, {Santos}, {Melo}, {Ammler-von
  Eiff}, {Torres}, {Quast}, {Gameiro}  \& {Sterzik}}{{Viana Almeida}
  et~al.}{2009}]{2009A&A...501..965V}
{Viana Almeida} P.,  {Santos} N.~C.,  {Melo} C.,  {Ammler-von Eiff} M.,
  {Torres} C.~A.~O.,  {Quast} G.~R.,  {Gameiro} J.~F.,   {Sterzik} M.,  2009,
  \mn@doi [\aap] {10.1051/0004-6361/200811194}, \href
  {https://ui.adsabs.harvard.edu/abs/2009A&A...501..965V} {501, 965}

\bibitem[\protect\citeauthoryear{{Wright}, {Drake}, {Drew}  \& {Vink}}{{Wright}
  et~al.}{2010}]{2010ApJ...713..871W}
{Wright} N.~J.,  {Drake} J.~J.,  {Drew} J.~E.,   {Vink} J.~S.,  2010, \mn@doi
  [\apj] {10.1088/0004-637X/713/2/871}, \href
  {https://ui.adsabs.harvard.edu/abs/2010ApJ...713..871W} {713, 871}

\bibitem[\protect\citeauthoryear{{Wright}, {Drew}  \& {Mohr-Smith}}{{Wright}
  et~al.}{2015}]{2015MNRAS.449..741W}
{Wright} N.~J.,  {Drew} J.~E.,   {Mohr-Smith} M.,  2015, \mn@doi [\mnras]
  {10.1093/mnras/stv323}, \href
  {https://ui.adsabs.harvard.edu/abs/2015MNRAS.449..741W} {449, 741}

\bibitem[\protect\citeauthoryear{{Zanni} \& {Ferreira}}{{Zanni} \&
  {Ferreira}}{2013}]{2013A&A...550A..99Z}
{Zanni} C.,  {Ferreira} J.,  2013, \mn@doi [\aap]
  {10.1051/0004-6361/201220168}, \href
  {https://ui.adsabs.harvard.edu/abs/2013A&A...550A..99Z} {550, A99}

\makeatother
\end{thebibliography}

\bsp
\label{lastpage}
\end{document}